\documentclass[twoside,twocolumn,9pt]{article}
\usepackage{extsizes}
\usepackage[super,sort&compress,comma]{natbib} 
\usepackage[version=3]{mhchem}
\usepackage[left=1.5cm, right=1.5cm, top=1.785cm, bottom=2.0cm]{geometry}
\usepackage{balance}
\usepackage{mathptmx}
\usepackage{sectsty}
\usepackage{graphicx} 
\usepackage{lastpage}
\usepackage[format=plain,justification=justified,singlelinecheck=false,font={stretch=1.125,small,sf},labelfont=bf,labelsep=space]{caption}
\usepackage{float}
\usepackage{fancyhdr}
\usepackage{fnpos}
\usepackage[english]{babel}
\addto{\captionsenglish}{%
  
}
\usepackage{array}
\usepackage{droidsans}
\usepackage{charter}
\usepackage[T1]{fontenc}
\usepackage[usenames,dvipsnames]{xcolor}
\usepackage{setspace}
\usepackage[compact]{titlesec}
\usepackage{hyperref}

\usepackage{epstopdf}

\definecolor{cream}{RGB}{222,217,201}

\graphicspath{{figures/}}
\usepackage{amsmath,bm}
\usepackage{amsbsy}

\begin{document}

\pagestyle{fancy}
\thispagestyle{plain}
\fancypagestyle{plain}{
\renewcommand{\headrulewidth}{0pt}
}

\makeFNbottom
\makeatletter
\renewcommand\LARGE{\@setfontsize\LARGE{15pt}{17}}
\renewcommand\Large{\@setfontsize\Large{12pt}{14}}
\renewcommand\large{\@setfontsize\large{10pt}{12}}
\renewcommand\footnotesize{\@setfontsize\footnotesize{7pt}{10}}
\makeatother

\renewcommand{\thefootnote}{\fnsymbol{footnote}}
\renewcommand\footnoterule{\vspace*{1pt}%
\color{cream}\hrule width 3.5in height 0.4pt \color{black}\vspace*{5pt}} 
\setcounter{secnumdepth}{5}

\makeatletter 
\renewcommand\@biblabel[1]{#1}            
\renewcommand\@makefntext[1]%
{\noindent\makebox[0pt][r]{\@thefnmark\,}#1}
\makeatother 
\renewcommand{\figurename}{\small{Fig.}~}
\sectionfont{\sffamily\Large}
\subsectionfont{\normalsize}
\subsubsectionfont{\bf}
\setstretch{1.125} 
\setlength{\skip\footins}{0.8cm}
\setlength{\footnotesep}{0.25cm}
\setlength{\jot}{10pt}
\titlespacing*{\section}{0pt}{4pt}{4pt}
\titlespacing*{\subsection}{0pt}{15pt}{1pt}

\fancyfoot{}
\fancyfoot[LO,RE]{\vspace{-7.1pt}\includegraphics[height=9pt]{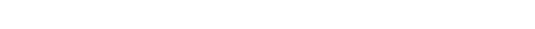}}
\fancyfoot[CO]{\vspace{-7.1pt}\hspace{13.2cm}\includegraphics{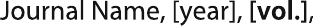}}
\fancyfoot[CE]{\vspace{-7.2pt}\hspace{-14.2cm}\includegraphics{head_foot/RF}}
\fancyfoot[RO]{\footnotesize{\sffamily{1--\pageref{LastPage} ~\textbar  \hspace{2pt}\thepage}}}
\fancyfoot[LE]{\footnotesize{\sffamily{\thepage~\textbar\hspace{3.45cm} 1--\pageref{LastPage}}}}
\fancyhead{}
\renewcommand{\headrulewidth}{0pt} 
\renewcommand{\footrulewidth}{0pt}
\setlength{\arrayrulewidth}{1pt}
\setlength{\columnsep}{6.5mm}
\setlength\bibsep{1pt}

\makeatletter 
\newlength{\figrulesep} 

\newcommand{\topfigrule}{\vspace*{-1pt}%
\noindent{\color{cream}\rule[-\figrulesep]{\columnwidth}{1.5pt}} }

\newcommand{\botfigrule}{\vspace*{-2pt}%
\noindent{\color{cream}\rule[\figrulesep]{\columnwidth}{1.5pt}} }

\newcommand{\dblfigrule}{\vspace*{-1pt}%
\noindent{\color{cream}\rule[-\figrulesep]{\mathrmwidth}{1.5pt}} }

\makeatother

\twocolumn[
  \begin{@twocolumnfalse}

\noindent\LARGE{{Onsager's variational principle in active soft matter}} \\
\vspace{0.3cm}  \vspace{0.3cm} \\ 
\noindent\large{Haiqin Wang\textit{$^{a,b}$}, Tiezheng Qian\textit{$^{c}$}, and Xinpeng Xu\textit{$^{b,a}$}$^{\ast}$} 
\\

\noindent\normalsize{ 
Onsager’s variational principle (OVP) was originally proposed by Lars Onsager in 1931 [L. Onsager, \emph{Phys. Rev.}, 1931, \textbf{37}, 405]. This fundamental principle provides a very powerful tool for formulating thermodynamically consistent models. It can also be employed to find approximate solutions, especially in the study of soft matter dynamics. In this work, OVP is extended and applied to the dynamic modeling of active soft matter such as suspensions of bacteria and aggregates of animal cells. We first extend the general formulation of OVP to active matter dynamics where active forces are included as external non-conservative forces. We then use OVP to analyze the directional motion of individual active units: a molecular motor walking on a stiff biofilament and a toy two-sphere microswimmer. Next we use OVP to formulate a diffuse-interface model for an active polar droplet on a solid substrate. In addition to the generalized hydrodynamic equations for active polar fluids in the bulk region, we have also derived thermodynamically consistent boundary conditions. Finally, we consider the dynamics of a thin active polar droplet under the lubrication approximation. We use OVP to derive a generalized thin film equation and then employ OVP as an approximation tool to find the spreading laws for the thin active polar droplet. By incorporating the activity of biological systems into OVP, we develop a general approach to construct thermodynamically consistent models for better understanding the emergent behaviors of individual animal cells and cell aggregates or tissues.}



 \end{@twocolumnfalse} \vspace{0.6cm}
]

\renewcommand*\rmdefault{bch}\normalfont\upshape
\rmfamily
\section*{}
\vspace{-1cm}



\footnotetext{\textit{$^{a}$~Technion -- Israel Institute of Technology, Haifa, 32000, Israel.}}
\footnotetext{\textit{$^{b}$~Physics Program, Guangdong Technion -- Israel Institute of Technology, Shantou, Guangdong 515063, China.}} 
\footnotetext{\textit{$^{c}$~Department of Mathematics, The Hong Kong University of Science and Technology, Clear Water Bay, Kowloon, Hong Kong, China.}}
\footnotetext{\textit{$^{\ast}$~Correspondence author, E-mail: xu.xinpeng@gtiit.edu.cn}}





\section{Introduction}\label{Introduction}

Active matter represents a novel type of nonequilibrium systems that contain a large number of self-propelling particles or creatures moving in fluids or more complex environments~\cite{Marchetti2013,Sam2013a,Bechinger2016,Prost2015,Menzel2015,Marenduzzo2015,Marenduzzo2016,Cates2015Rev,Lauga2009,Elgeti2015}. The self-propelling units are considered to be active in the sense that they are capable of continuously converting fuel or chemical energy (stored internally or in the ambient) into directional motion or mechanical work. \textcolor{black}{Active matter containing self-propelled units is ubiquitous in biology and in many artificial systems.} Examples in biology are abundant and occur at all length scales, ranging from bacteria suspensions~\cite{Gachelin2014,Lauga2009,Elgeti2015,Zhang2010,Bechinger2016} to animal cells~\cite{Sam2013a,Prost2015}, 
animal cell aggregates (or tissues)~\cite{Joanny2010,Len2013,Len2020,Julicher2017,Marchetti2019}, bird and fish flocks~\cite{Toner1995,Hemelrijk2012}, 
and \textcolor{black}{pedestrian crowds}~\cite{Helbing2001,Castellano2009}. 
Artificially made active matter~\cite{Bechinger2016} includes layers of vibrated granular rods~\cite{Marchetti2013,Aranson2007}, collections of robots~\cite{Scholz2018}, and suspensions of colloidal or nanoscale particles~\cite{Zhang2017,He2018,YangNi2020} that are propelled through catalytic activities at their surfaces. 

A distinctive feature of active matter is that the system is locally driven out of equilibrium by active units at the length scale of a constituent component~\cite{Marchetti2013,Sam2013a,Joanny2015,Bechinger2016}. This is distinct from those nonequilibrium systems that are driven at the system boundaries~\cite{Hohenberg1993}. 
The presence of \textcolor{black}{self-propelled units in active matter} breaks the detailed balance and time-reversal symmetry (TRS)~\cite{Marchetti2013,Menzel2015,Cates2015Rev,Bechinger2016}, resulting in a wealth of intriguing macroscopic structures and behaviors, such as spontaneous flows~\cite{Prost2015,Bechinger2016}, \textcolor{black}{motility-induced phase separation}~\cite{Marenduzzo2015,Cates2015Rev,Bechinger2016}, unusual mechanical and rheological properties~\cite{Sam2013a,Elgeti2015}, 
wave propagation and sustained oscillations even in the absence of inertia~\cite{Gerisch2004,Weiner2007,Inagaki2017}, \emph{etc}. One of the most interesting questions in the nonequilibrium dynamics of active matter is how the local driving forces operating at the small scale of individual active unit can produce the observable macroscopic emergent phenomena at the large scale of the whole system. Answering this question will not only shed new light on the fundamental statistical mechanics~\cite{Marenduzzo2015,Cates2015Rev,Bechinger2016,YangNi2020}, but also deepen our understanding of biological processes~\cite{Marchetti2013,Sam2013a,Prost2015,Bechinger2016}, \textcolor{black}{and help design new generations of biomimetic active materials that balance structural flexibility and stability}~\cite{Marchetti2013,Menzel2015,Bechinger2016}.

The study of active matter can be brought into the framework of condensed matter physics based on the consideration that the collective behaviors of active matter emerge from the interactions among the constituent self-propelling units 
and the dissipation mechanisms operating inside the system. In particular, soft condensed matter physics provides \textcolor{black}{many} useful model systems for reference to active matter~\cite{Sam2013a,Menzel2015}, \emph{e.g.} the wetting of substrates by liquid droplets~\cite{Joanny1992,DeGennes1985,Bonn2009}, 
the dynamics of colloid suspensions~\cite{Tanaka2012,Winkler2014}, the dynamics of nematic liquid crystals~\cite{deGennes1993}, the dynamics and rheology of polymer gels~\cite{Ross1994,Doi2009}, and the phase segregation of surfactants~\cite{Cates1990,Sam1991}, \emph{etc}. The major challenge is to couple these model systems with active and molecularly specific processes, such as the active force generation by self-propelling units, and the binding and unbinding of transmembrane adhesion receptors on solid substrates~\cite{Sam2013a,Menzel2015}. Over the last decade, several soft matter systems have been revisited with a focus on this point of view~\cite{Sam2013a}. The physical understanding for the emergent structures and behaviors of active soft matter has been rapidly growing~\cite{Marchetti2013,Sam2013a,Bechinger2016,Prost2015,Menzel2015,Marenduzzo2015,Marenduzzo2016,Cates2015Rev,Lauga2009,Elgeti2015}, with particular attention paid to dry active matter~\cite{Toner1995,Bechinger2016}, active polar fluids~\cite{Baskaran2010}, active nematics~\cite{Cates2018}, active gels~\cite{Ramaswamy2002}, and active membranes~\cite{Prost1996}. 

Theoretically there have been two major approaches to the study of active soft matter: particle-based models~\cite{Marenduzzo2015,Marenduzzo2016,Cates2015Rev,YangNi2020,Kafri2019,Bechinger2016} and continuum phenomenological models~\cite{Marchetti2013,Sam2013a,Prost2015,Menzel2015,Marenduzzo2015,Cates2018,Len2013,Len2020}. In particle-based models, the active units are usually modeled as self-propelled particles with fixed or variable speed and random orientation 
moving in an inert background, following the seminal work of Vicsek \emph{et al.}~\cite{vicsek1995}. It provides a straightforward approach to the study of active soft matter 
with an emphasis on the order and fluctuations rather than the forces and mechanics~\cite{Marenduzzo2016,Cates2015Rev,Bechinger2016,YangNi2020}. In continuum phenomenological models, active units are represented by a smooth density field rather than individually resolved particles. A continuum model for active soft matter is usually constructed by modifying the dynamic model of a proper reference soft matter system~\cite{Marchetti2013,Sam2013a,Prost2015,Menzel2015,Marenduzzo2015,Cates2018}. This is typically accomplished by adding a minimal set of extra terms that cannot be derived from any free energy or dissipation functions. This is an effective way to introduce the activity and break the TRS such as active forces, active fluxes, 
and active chemical potentials~\cite{Marchetti2013,Cates2014,Cates2015,Cates2018}. However, there is another more systematic way of including activity by introducing the  mechanochemical coupling between passive dissipative processes and some relevant biochemical reactions~\cite{Marchetti2013,Prost2015} 
in Onsager's framework of irreversible thermodynamics~\cite{deGroot1984,Gyarmati1970}. The two theoretical approaches are complementary. The particle-based approach involves only a small number of parameters for each active unit, and therefore the theoretical predictions can be readily compared with experiments for some model active systems such as self-propelled colloids~\cite{ChateZhang2019}. 
However, the model for interacting self-propelled particles sometimes oversimplifies the problem, and hence may lose some generality and applicability of its conclusions when applied to real systems, especially \emph{in vivo} biological systems~\cite{Marchetti2013,Sam2013a,Prost2015,Cates2018}. 
By contrast, the formulation of phenomenological models is based upon symmetry consideration, conservation laws of mass, momentum, and angular momentum, and laws of thermodynamics. 
This gives the continuum approach a large range of applicability and generality when applied to real biological processes~\cite{Marchetti2013,Sam2013a,Prost2015,Cates2018}. 
In this work, we focus on the continuum phenomenological models and show that Onsager’s variational principle, which has been widely used in the study of soft matter dynamics~\cite{Doi2013soft}, can be extended for the study of active soft matter.  

Onsager's variational principle (OVP) was originally proposed by Lars Onsager in his seminar papers in 1931~\cite{Onsager1931a,Onsager1931b}. 
He showed that for irreversible processes in a near-equilibrium thermodynamic system, the thermodynamic fluxes can be written as linear combinations of conjugate thermodynamic forces, and the proportionality coefficient matrix must be positive-definite and symmetric according to Onsager's reciprocal relations (ORR). The ORR lay the foundation for the theoretical framework of linear irreversible thermodynamics. 
In the end of his paper, Onsager proposed OVP as a variational principle that is equivalent to the linear force-flux relations in describing dissipative dynamics. In addition, OVP can be regarded as an extension of ``the principle of the least dissipation of energy'' proposed by Lord Rayleigh~\cite{Rayleigh1873}. For isothermal systems, OVP takes a simple form as follows~\cite{Onsager1945,Qian2006,Doi2013soft}. The irreversible processes described by the thermodynamic fluxes $\dot{\bm{\alpha}}$ follow the dynamic path that minimizes the function:
\begin{equation}\label{eq:introduction-Ray} 
{\cal R}(\dot{\bm{\alpha}})= \Phi(\dot{\bm{\alpha}})+\dot{F} (\dot{\bm{\alpha}};\bm{\alpha}).
\end{equation}
Here the function ${\cal R}$ is called the \emph{Rayleighian} as suggested by \textcolor{black}{Doi and Edwards}~\cite{DoiEdwards1986}, 
$\Phi$ is the dissipation function which is quadratic in $\dot{\bm{\alpha}}$ when the system is close to equilibrium, 
and $\dot{F}$ is the rate of change of the free energy in the isothermal system. Onsager later used his principle to study the diffusion in electrolyte solutions~\cite{Onsager1945}. 
However, OVP has not been widely recognized and applied to describe irreversible processes for a long time until 1953 when 
Onsager and Machlup~\cite{Onsager1953} established the statistical mechanical foundation of OVP.
Since then, OVP and its relationship with other thermodynamic variational principles have been extensively studied and identified~\cite{ono1961}. 
In recent years, OVP has been widely used as an indispensable and powerful tool for the study of nonlinear and nonequilibrium phenomena of soft matter~\cite{Qian2006,Doi2011,Doi2013soft,Doi2020,ChunLiu2020,QWang2020}. 

OVP can be used to derive many transport equations for soft matter dynamics~\cite{Doi2013soft,Doi2011},
\emph{e.g.}, the Stokes equation for incompressible low-Reynolds-number flows~\cite{Qian2006}, 
the diffusion equation~\cite{Doi2011,Doi2013soft}, 
the reaction-diffusion equations for (low molecular weight) multi-component solutions~\cite{ChunLiu2020,Gyarmati1970,QWang2020}, 
the thin film evolution equations~\cite{Xu2015,DoiDi2018,Doi2019,Doi2020}, 
the phase field model for two-phase hydrodynamics~\cite{Qian2006,xuQian2017}, 
the electrorheological hydrodynamic equation~\cite{sheng2008onsager}, 
the two-fluid model for the phase separation dynamics in colloids and polymers~\cite{Doi2013soft,Doi2011,Doi2020}, the dynamics of polymer gels~\cite{Doi2009,Doi2020}, 
the dynamic equations of lipid membrane~\cite{Krishnan2016,Oya2018,Arroyo2018}, \emph{etc}. Moreover, OVP also provides a very convenient way to derive thermodynamically consistent 
boundary conditions that supplement the transport equations in the bulk region~\cite{Qian2006,xuQian2017}. 
Examples include 
the generalized Navier boundary condition (GNBC) for the contact line hydrodynamics~\cite{Qian2006},
the generalized nemato-hydrodynamic boundary conditions for liquid crystals~\cite{Qian2008}, 
and the boundary conditions for block copolymer solution films~\cite{xu2019defect}, \emph{etc}. 
In addition, Doi and his collaborators have recently proposed that OVP can be used as 
a direct variational method to find approximate solutions for complex soft matter dynamics~\cite{Doi2015,Doi2019,Doi2020}. 
This approximation method has been successfully used to study the evolution of droplets and 
thin films~\cite{Doi2016,Doi2017,DoiDi2018,Doi2020}, 
the dynamics of the beads-on-string structure of viscoelastic polymer filaments~\cite{Zhou2018,Doi2020}, 
the sedimentation in colloidal suspensions~\cite{Sui2018}, 
and the translocation of a vesicle through a narrow hole~\cite{khunpetch2018translocation}, \emph{etc}. 
Although OVP has been widely applied with great successes in the study of inert soft matter dynamics, 
it is rarely used in the study of active soft matter dynamics~\cite{Vernerey2016,Tu2020,QWang2016,QWang2020}. 
In the present work, we will show that OVP can be readily extended to include biochemical activity and conveniently applied to study the emergent structures and behaviors of active soft matter. 
OVP can not only be applied to formulate thermodynamically consistent models, but also be used to generate approximate solutions for the complex dynamics of active soft matter. 

This paper is organized as follows. In Sec.~\ref{Sec:OVP}, a brief review of OVP is provided, 
and a simple extension of OVP is presented for applications to active matter, in which the active forces are treated as non-conservative forces that cannot be derived from any free energy and dissipation functions. In the next three sections, we apply OVP and its extended form to three representative 
active matter problems motivated by the biology of bacteria and animal cells. In Sec.~\ref{Sec:Appl-ActiveUnit}, we present the first application of OVP to the directional motion of 
an individual active unit, \emph{e.g.}, 
a molecular motor walking on a stiff biofilament and a toy two-sphere microswimmer moving in a viscous fluid. 
In Sec.~\ref{Sec:Appl-ActivePolar}, we consider the two-phase hydrodynamics for active polar droplets. 
We use OVP to formulate a diffuse-interface model 
for an active polar droplet on a solid substrate. 
This hydrodynamic model is thermodynamically consistent and consists of hydrodynamic equations in the bulk region and boundary conditions at the solid surface. 
In Sec.~\ref{Sec:Appl-DropSpread}, we consider a thin active polar droplet moving on a solid substrate in two dimensions. Under the lubrication approximation, we firstly apply OVP to derive the classical thin film equation that has been obtained previously. We then use OVP as an approximation tool to find the scaling laws for the spreading of a thin active droplet in the respective limits of negligible activity and strong activity. In Sec.~\ref{Sec:Conc}, we summarize our major results, make some general remarks, and envision a few potential applications of OVP 
to more realistic biological problems.

\par 
 

\section{Variational principles for active matter dynamics}
\label{Sec:OVP}

In this section, we show that the original variational principle of Onsager~\cite{Onsager1931a,Onsager1953} can be easily modified to study the dynamics of active soft matter by including active forces as non-conservative forces that can not be derived from any free energy function. Since we are mostly interested in the flow, diffusion, and biochemical reactions in active soft matter, we, therefore, limit our discussions to isothermal systems where temperature is assumed to be constant. 
\subsection{Onsager's variational principle and Onsager-Machlup variational principle}\label{Sec:OVP-OVPOMVP}

Consider a non-equilibrium isothermal system that is characterized by a set of coarse-grained, slow state variables, $\bm{\alpha} \equiv \{\alpha_1, \, \alpha_2, \,\dots, \, \alpha_N\}$. The dynamics of the system is then described by the time evolution of $\bm{\alpha}(t)$ which is governed, in the linear response regime near equilibrium, by the general (overdamped) Langevin equation~\cite{Onsager1953}
\begin{equation}\label{eq:OVP-Langevin}
\sum_{j=1}^{N} \zeta_{ij} \dot{\alpha}_{j} =f_{i}(\bm{\alpha},t)+f_{r i}(t),
\end{equation}
in which $\bm{\zeta}=\{\zeta_{ij}\}$ is the friction coefficient matrix that generally depends on the state variables~\cite{QWang2020} $\bm{\alpha}$. Here we assume that all the state variables $\bm{\alpha}$ have the same time parity, and therefore, according to Onsager's reciprocal symmetry, the friction matrix $\bm{\zeta}$ is not only 
positive definite (due to the second law of thermodynamics) 
but also symmetric~\cite{Onsager1931a}, {\emph{i.e.}}, $\zeta_{ij}=\zeta_{ji}$. State variables with different time parities have been discussed briefly in the appendix Sec.~\ref{Sec:AppA-TimeParity}. The stochastic force $f_{r}(t)$ is assumed to be an uncorrelated white noise  
\begin{equation}\label{eq:OVP-fr}
\left\langle f_{r i}(t)\right\rangle=0, \quad\left\langle f_{r i}(t) f_{r j}\left(t^{\prime}\right)\right\rangle=2 \zeta_{i j} k_{B} T \delta\left(t-t^{\prime}\right),
\end{equation}
with $k_B$ denoting the Boltzmann constant and $T$ the temperature. The generalized force $f_{i}$, in general, includes two types of forces: $f_{i}(\bm{\alpha},t)=f_{ci}(\bm{\alpha})+f_{ai}(\bm{\alpha},t)$, in which the conservative force $f_{ci}(\bm{\alpha})$ can be derived from some free energy $F(\bm{\alpha})$ by 
\begin{equation}\label{eq:OVP-fc}
f_{c i}(\bm{\alpha})=-\frac{\partial F(\bm{\alpha})}{\partial \alpha_{i}},
\end{equation} 
and $f_{ai}(\bm{\alpha},t)$ is the \emph{active force}, a non-conservative force that cannot be derived from any energy function. Physically, the active forces arise from the persistent consumption of chemical energy and they continuously drive the system out of equilibrium locally at the small scale of individual active unit. For example, the active forces can be generated by biochemical reactions such as ATP hydrolysis in animal cells/tissues~\cite{Marchetti2013,Prost2015}, or by external fields such as light acting on active colloids with photosensitive coatings~\cite{Bechinger2016,Marenduzzo2016}. 
The active forces also break the time-reversal symmetry (TRS) of the system (in a sense different from the breakdown of TRS due to friction)~\cite{Cates2018,Cates2014,Cates2015b,Cates2015Rev}. For example, the self-propelling force in active colloids drives the persistent motion of colloids in some direction, resulting in the intrinsic breakdown of TRS \cite{Cates2018}. 

From the Langevin equation (\ref{eq:OVP-Langevin}), we calculate the transition probability~\cite{Onsager1953} $P\left(\bm{\alpha}^{\prime}, t+dt\mid \bm{\alpha}, t\right)$ from the state $\bm{\alpha}$ at time $t$ to $\bm{\alpha}^{\prime}$ at $t+dt$:
\begin{align}\label{eq:OVP-Pshort}
&P\left(\bm{\alpha}^{\prime}, t+dt\mid \bm{\alpha}, t\right)=\frac{1}{\sqrt{(2\pi)^N \det(2\bm{\zeta}^{-1}k_BT)dt}} \exp\left[-\frac{{\cal O}(\bm{\alpha}^{\prime},dt,\bm{\alpha},t)}{2k_BT} \right]
\end{align} 
for $\bm{\alpha}^{\prime}$ close to $\bm{\alpha}$, and $dt$ is an infinitesimal time interval. Here the Onsager-Machlup function $\cal O$ is defined by
\begin{align}\label{eq:OVP-Oshort}
{\cal O} &= \frac{1}{2dt}\zeta_{ij}\left[\alpha_i'-\alpha_i -\mu_{ik}f_k (\bm{\alpha},t)dt\right]\left[\alpha_j'-\alpha_j -\mu_{jk}f_k (\bm{\alpha},t)dt\right] \nonumber \\ 
&= \frac{1}{2}\zeta_{ij}\left[\dot{\alpha}_i -\mu_{ik}f_k (\bm{\alpha},t)\right]\left[\dot{\alpha}_j -\mu_{jk}f_k (\bm{\alpha},t)\right]dt
\end{align}
with the rates $\dot{\alpha}_i\equiv (\alpha_i'-\alpha_i)/dt$ and $\mu_{ij}$ being the mobility coefficient matrix that is positive definite and symmetric, and satisfies $\zeta_{ij}\mu_{jk}=\delta_{ik}$. From the transition probability $P\left(\bm{\alpha}^{\prime}, t+dt\mid \bm{\alpha}, t\right)$ in Eq.~(\ref{eq:OVP-Pshort}), the most probable transition occurring between $\bm{\alpha}$ and nearby $\bm{\alpha}'$ is the one which minimizes the Onsager-Machlup function $\cal O$ or equivalently the function
\begin{equation}\label{eq:OVP-Ray} 
{\cal R}(\dot{\bm{\alpha}};\bm{\alpha}) = \Phi (\dot{\bm{\alpha}},\dot{\bm{\alpha}})+\dot{F} (\dot{\bm{\alpha}};\bm{\alpha}) -\dot{W}_a(\dot{\bm{\alpha}};\bm{\alpha}) 
\end{equation}
with respect to $\bm{\alpha}^{\prime}$ or equivalently, the rates $\dot{\bm{\alpha}}$ for prescribed $\bm{\alpha}$. Here ${\cal R}$ is called the \emph{Rayleighian}~\cite{Doi2011,Doi2020}, $\Phi (\dot{\bm{\alpha}},\dot{\bm{\alpha}})\equiv \frac{1}{2}\zeta_{ij}\dot{\alpha}_i\dot{\alpha}_j$ is the positive-definite dissipation-function, $\dot{F}(\dot{\bm{\alpha}};\bm{\alpha}) \equiv\frac{\partial F}{\partial \alpha_i}\dot{\alpha}_i=-f_{ci}(\bm{\alpha},t)\dot{\alpha}_i$ is the change rate of free energy, and $\dot{W}_a(\dot{\bm{\alpha}};\bm{\alpha}) =f_{ai}(\bm{\alpha},t)\dot{\alpha}_i$ is the work power done by the active force, $\bm{f}_{a}$. The Euler–Lagrange equation for minimizing ${\cal R}$ in Eq.~(\ref{eq:OVP-Ray}) with respect to $\dot{\bm{\alpha}}$ is 
\begin{equation}\label{eq:OVP-DynEqn}
\zeta_{ij}\dot{\alpha}_j=f_{i}(\bm{\alpha},t)=-\frac{\partial F(\bm{\alpha})}{\partial \alpha_{i}}+f_{ai}(\bm{\alpha},t),
\end{equation}
as expected from the Langevin equation (\ref{eq:OVP-Langevin}) for dynamic processes where the stochastic forces and fluctuations can be neglected\footnote[1]{It is interesting to mention that in active matter, the presence of active forces and the breakdown of TRS can significantly change the statistical behaviors of the system such as the distribution of $\bm{\alpha}$ and barrier crossing kinetics. The dynamics of downhill and uphill processes are very different even for passive systems, and hence must be treated using their respective variational approaches. In this work, we are interested in the active dynamics that would reduce to downhill processes if the activity vanishes. However, the fluctuation effects can still be investigated in a variational framework~\cite{Kafri2019, Qian2020}.}.
\textcolor{black}{The dynamic equations in Eq.~(\ref{eq:OVP-DynEqn}) state the balance among the dissipative (frictional) forces, 
the conservative forces derived from the free energy, and the active forces that are non-conservative and generated by 
biochemical reactions. This variational principle of minimizing the Rayleighian ${\cal R}$ with respect to the rates  $\dot{\bm{\alpha}}$  was originally proposed by Lars Onsager in 1931 \cite{Onsager1931a,Onsager1953} and usually referred to as \emph{The principle of least dissipation of energy} or simply \emph{Onsager's variational principle} (OVP). }

\textcolor{black}{We would like to give some remarks on OVP as follows.} 

\textcolor{black}{(i) A term $\Psi(\bm{f}(\bm{\alpha},t),\bm{f}(\bm{\alpha},t))\equiv\frac{1}{2}\mu_{ik}f_{i}(\bm{\alpha},t)f_{k}(\bm{\alpha},t)$ appearing in the Onsager-Machlup function ${\cal O}$ in Eq.~(\ref{eq:OVP-Oshort}) does not contribute to the Rayleighian ${\cal R}$ because we are considering the most probable state $\bm{\alpha}'$ at an immediate future time close to $t$, 
which is to be determined from the prescribed state $\bm{\alpha}$ and hence the prescribed forces $\bm{f}(\bm{\alpha},t)$ and $\Psi(\bm{f}(\bm{\alpha},t),\bm{f}(\bm{\alpha},t))$. } 

\textcolor{black}{(ii) The active matter may also be subject to the influence of some external forces 
that do not arise locally from the consumption of chemical energy of the system. This can be treated by subtracting 
the work power $\dot{W}_{\mathrm{ext}}(\dot{\bm{\alpha}};\bm{\alpha})$ done by the external forces from 
the Rayleighian ${\cal R}$ in Eq.~(\ref{eq:OVP-Ray}): 
\begin{equation}\label{eq:OVP-RayExt} 
{\cal R}(\dot{\bm{\alpha}};\bm{\alpha}) = \Phi (\dot{\bm{\alpha}},\dot{\bm{\alpha}})+\dot{F} (\dot{\bm{\alpha}};\bm{\alpha}) -\dot{W}_a(\dot{\bm{\alpha}};\bm{\alpha}) -\dot{W}_{\mathrm{ext}}(\dot{\bm{\alpha}};\bm{\alpha}).
\end{equation}
Minimization of ${\cal R}$ leads to an Euler-Lagrange equation that includes the external forces, in a form generalized from Eq.~(\ref{eq:OVP-DynEqn}). }

\textcolor{black}{(iii) In a continuum model of active matter, 
the set of slow variables $\bm{\alpha}$ can represent the field variables both in the bulk and at the boundary. 
Then Eq.~(\ref{eq:OVP-DynEqn}) derived from OVP gives the dynamic equations both in the bulk and at the boundary, 
with the latter becoming dynamic boundary conditions. 
Furthermore, if there are external forces applied at the system boundary, then their contributions may be described by $\dot{W}_{\mathrm{ext}}$ in the Rayleighian in Eq.~(\ref{eq:OVP-RayExt}).}


\textcolor{black}{(iv) It is important to note that OVP is a local principle that can be used to find the most probable state 
only in the immediate future (without additional constraints).} To locate the most probable paths that can take the system to the far future under various constraints~\cite{Onsager1953,Ren2020}, we can divide the long time interval (say, $t-t_0$) into $N_t$ sub-intervals with $dt=(t-t_0)/N_t$, then
\begin{align}\label{eq:OVP-Plong}
P\left(\bm{\alpha}, t \mid \bm{\alpha}_0,t_0\right)&=\lim_{N_t\to \infty}\left[\prod_{m=1}^{N_t-1}\int d\bm{\alpha} \prod_{m=0}^{N_t-1} P\left(\bm{\alpha}_{m+1}, t_{m+1}\mid\bm{\alpha}_m,t_m\right)\right] \\ \nonumber
&\propto \int {\cal D} \bm{\alpha}(t) \exp\left\{-\frac{{\cal O}[\bm{\alpha}(t)]}{2k_BT}\right\},
\end{align} 
where ${\cal O}[\bm{\alpha}(t)]$ is called the Onsager-Machlup functional or integral defined by
\begin{equation}\label{eq:OVP-Olong}
{\cal O}\equiv \int_{t_0}^{t} d\tau \left[\frac{1}{2}\zeta_{ij}(\dot{\alpha}_i-\mu_{ik}f_k (\bm{\alpha},t))(\dot{\alpha}_j-\mu_{jk}f_k (\bm{\alpha},t)) \right],
\end{equation}  
or equivalently, into the following quadratic form
\begin{equation}\label{eq:OVP-Olong-LeastSquare}
{\cal O}\equiv \int_{t_0}^{t} d\tau \left[\frac{1}{2}\zeta_{ij}(\dot{\alpha}_i- \dot{\alpha}_i^*)(\dot{\alpha}_j-\dot{\alpha}_j^*) \right]
\end{equation}  
with $\dot{\alpha}_i^*(\bm{\alpha},t) = \mu_{ik}(\bm{\alpha})f_k(\bm{\alpha},t)$ being the actual rates of the system at state $\bm{\alpha}$ and time $t$. Note that $\dot{\bm{\alpha}}$ is defined by the time derivative of the state variables ${\bm{\alpha}}$, while $\dot{\bm{\alpha}}^*$ is defined by the force $\bm{f}(\bm{\alpha},t)$ and depends only on the state variable $\bm{\alpha}$ and time $t$. The Onsager-Machlup integral is non-negative definite and equals to zero (the minimum) only when $\dot{\bm{\alpha}}$ equals to the actual kinetic path, {\emph{i.e.}, $\dot{\bm{\alpha}}=\dot{\bm{\alpha}}^*$}. Hence the variational principle can be stated that nature chooses the kinetic path which minimizes the Onsager-Machlup functional ${\cal O}[\bm{\alpha}(t)]$ with respect to the state variables $\bm{\alpha}(t)$. This variational principle is called \emph{Onsager-Machlup variational principle} (OMVP)~\cite{Doi2019,Doi2020}. In contrast to OVP, a local principle that predicts the most probable state in the immediate future, the OMVP is a global principle that can determine the most probable path taking the system to the far future under various constraints. For example, it can be used to determine the long-time behaviors such as the steady-states~\cite{Doi2019,Doi2020}. It can also be used to locate the most probable transition pathway that takes the system from one free energy minimum to another~\cite{Touchette2009,Ren2020}.

\subsection{Direct variational methods for approximation solutions}

The variational principles introduced previously can be used to obtain thermodynamically-consistent dynamic equations as well as matching boundary conditions. However, the \textcolor{black}{resulting} equation systems are usually difficult to solve both analytically and numerically. This subsection deals with some direct variational methods of finding approximate solutions that are obtained directly from above variational principles for the system dynamics in both short and long time scales~\cite{Doi2015,Doi2019,Doi2020,Reddy2017}. 

Variational principles have been proposed in various fields of physics and several variational methods have been developed accordingly to find approximation solutions such as Ritz method and the least-squares method~\cite{Reddy2017}. Recently, Doi developed a Ritz-type variational method based on OVP~\cite{Doi2015} by assuming that state variables $\bm{\alpha}(t)$ is a certain function of a small number of parameters denoted by $\bm{a}=\left(a_{1}, \, a_{2}, \ldots \right)$, \emph{i.e.}, $\bm{\alpha}(t)=\bm{\alpha}(\bm{a}(t))$. Then the rate of the state variables $\dot{\bm{\alpha}}$ can be written as 
\begin{equation}
\dot{\alpha}_{i}=\sum_{m} \frac{\partial \alpha_{i}}{\partial a_{m}} \dot{a}_{m}
\end{equation}
and the Rayleighian is then a function of the rates of parameters $\mathcal{R}(\dot{\bm{a}})$. The dynamics of the system described by the temporal evolution of the parameters $\bm{a}(t)$ is determined by minimizing $\mathcal{R}$ with respect to $\dot{\bm{a}}$. 

Similarly, direct variational method can also be developed based on the OMVP to approximate the long time kinetic paths or states of the system~\cite{Doi2019}. We consider certain kinetic path which involves a parameter set $\bm{\alpha}(\bm{a}(t))$. The best guess for the actual path is the path which gives the smallest value of the Onsager-Machlup functional ${\cal O}[\bm{\alpha}(\bm{a}(t))]$ with respect to the parameter functions $\bm{a}(t)$. This variational method is similar to the least-square method but with a target function that is more physically meaningful based on physical principles~\cite{Doi2019,Reddy2017}.  

These direct variational methods are useful particularly when we have an idea for the probable kinetic path and can write down the functions $\bm{\alpha}(\bm{a}(t))$. It has been applied successfully to many problems in soft matter dynamics. In this work, we will show that these approximation methods can also be used to study the dynamics of active soft matter. 


\subsection{Advantages of variational approaches} 

The above variational principles are equivalent to Onsager’s kinetic equations in Eq.~(\ref{eq:OVP-DynEqn}) with kinetic coefficients satisfying Onsager's reciprocal relations, but these variational principles and the direct variational methods have several advantages to investigate the dynamics of non-equilibrium systems as follows.
\begin{itemize}
\item \emph{Scalar formulation}. The variational principles involve only physical quantities that can be defined without reference to a particular set of generalized coordinates, namely the dissipation function, free energy, and active work power. This formulation is therefore automatically invariant with respect to the choice of coordinates for the system, which allows us a great flexibility in choosing state variables and rates. 

\item \emph{Thermodynamic consistency}. The variational principles incorporate the intrinsic structure of Onsager’s theory of non-equilibrium thermodynamics clearly. They provide compact invariant ways of obtaining thermodynamically-consistent dynamic equation systems where the pairs of rates and forces are obtained automatically.  
 
\item \emph{Direct variational approximation tools}. The direct Ritz-type variational method of finding approximation solutions for the system dynamics bypasses the derivation of the Euler-Lagrange equations and goes directly from a variational statement of the problem to the solution of the Euler-Lagrange equations. This approximation method helps to pick up the most important dynamic behaviors and to simplify the calculations significantly from complicated partial differential equation systems to simple ordinary differential equations. In addition, the direct least-square-type variational method based on OMVP of minimizing the Onsager-Machlup integral further optimizes the search for more realistic kinetic paths and provides a new method of studying long-time steady-state dynamics of the system.
\end{itemize}

The above variational principles have been successfully applied to the diffusion in electrolyte solutions by Onsager himself in 1940s~\cite{Onsager1945}, and more recently applied to various soft matter systems such as multiphase flows~\cite{Qian2006,xuQian2017}, electrorheological fluids~\cite{sheng2008onsager}, colloid suspensions~\cite{Sui2018}, polymer solutions~\cite{Doi2013soft,Doi2011,Doi2020}, polymer gels~\cite{Doi2009,Doi2020}, liquid crystals~\cite{Doi2011,Qian2008}, vesicles~\cite{khunpetch2018translocation}, membranes~\cite{Krishnan2016,Oya2018,Arroyo2018} and so on. This indicates that OVP is an important principle in soft matter dynamics~\cite{Doi2011,Doi2019,Doi2020}. In this work, we present its applications to active soft matter dynamics that is mostly motivated by biological applications.

\textcolor{black}{Before ending this section, we would like to summarize the general steps for applying OVP 
to the dynamics of active soft matter for which the dynamic equations are not yet known or still controversial~\cite{Doi2011,Doi2020}.}

\textcolor{black}{(i) \emph{Choose a set of coarse-grained, slow  variables}, $\bm{\alpha} \equiv \{\alpha_1, \, \alpha_2, \,\dots \}$, to describe the time evolution of the macroscopic state of the system.}

\textcolor{black}{(ii) \emph{Construct the free energy function}, $F(\bm{\alpha})$, and calculate the rate of change of the free energy, $\dot{F} (\dot{\bm{\alpha}};\bm{\alpha})$.}

\textcolor{black}{(iii) \emph{Construct the dissipation function}, $\Phi (\dot{\bm{\alpha}},\dot{\bm{\alpha}})$, which is quadratic in the rates/fluxes $\dot{\bm{\alpha}}$.}

\textcolor{black}{(iv) \emph{Find the work power} done by the active forces, $\dot{W}_a(\dot{\bm{\alpha}};\bm{\alpha})$, based on the specific activity considered, and find the work power done by some other external forces, $\dot{W}_{\mathrm{ext}}(\dot{\bm{\alpha}};\bm{\alpha})$. The external forces are usually applied at the system boundary and do not arise locally from the consumption of chemical energy of the system.}

\textcolor{black}{(v) \emph{Minimize the Rayleighian} in Eq.~(\ref{eq:OVP-RayExt}): ${\cal R}(\dot{\bm{\alpha}};\bm{\alpha}) = \Phi (\dot{\bm{\alpha}},\dot{\bm{\alpha}})+\dot{F} (\dot{\bm{\alpha}};\bm{\alpha}) -\dot{W}_a(\dot{\bm{\alpha}};\bm{\alpha}) -\dot{W}_{\mathrm{ext}}(\dot{\bm{\alpha}};\bm{\alpha})$, with respect to the rates/fluxes $\dot{\bm{\alpha}}$. Note that some additional constraints on the system dynamics may need to be imposed by using Lagrange multipliers.} 

\noindent\textcolor{black}{Furthermore, if we have an idea about the most probable kinetic path, 
then we can write down the slow variables $\bm{\alpha}=\bm{\alpha}(\bm{a}(t))$ 
as functions of a small number of parameters, $\bm{a}=\left(a_{1}, \, a_{2}, \ldots \right)$. 
We can follow the above steps and obtain the Rayleighian as a function of $\dot{\bm{a}}$ and $\bm{a}$ 
as ${\cal R}(\dot{\bm{a}};\bm{a})$. The minimization of ${\cal R}$ with respect to $\dot{\bm{a}}$ 
will then provide an approximate description for the active matter dynamics directly.}

\section{Applications 1: Directional motion of individual active units}\label{Sec:Appl-ActiveUnit}

Activity in biology or in some artificial active systems usually arises from the consumption of ATP and the  mechanochemical cross-coupling, but sometimes activity appears simply as time-dependent constraints in geometric shapes. In this section, we consider the first applications of OVP to the directional motion of an individual active unit as shown in Fig.~\ref{Fig:ActiveUnits}: a molecular motor walking on a polar biofilament and a toy two-sphere microswimmer swimming in a viscous fluid.

\subsection{Molecular motors walking on biofilaments: mechano-chemical cross-coupling} \label{Sec:Appl-ActiveUnit1}

Molecular motor proteins are enzymes that bind adenosine triphosphate (ATP) and catalyze its hydrolysis to adenosine diphosphate (ADP) and inorganic phosphate ($\mathrm{P_i}$)~\cite{Alberts2007}:
\begin{equation}\label{eq:Appl-ActiveUnit-ATP} 
{\mathrm{ATP}}\rightleftharpoons {\mathrm{ADP}}+{\mathrm{P_i}}. 
\end{equation}
The chemical energy released from this ATP hydrolysis is partially converted into mechanical work or directional motion of motors along some stiff biofilaments that are made of other proteins. Animal cells \emph{in vivo} contain over a hundred different motor proteins, which can be classified into three different families: myosins moving along actin filaments, kinesins and dyneins moving along tubulin filaments. The underlying biofilaments are usually periodic and fairly rigid structures with a period $\xi \sim 10\, \mathrm{nm}$. They are moreover polar or asymmetric, so that one can define a ``plus'' end and a ``minus'' end. A given motor always moves in the same deterministic direction: myosin moves along actin filaments towards their plus end, and kinesins and dyneins move along tubulin filaments towards their plus and minus ends, respectively. Motor molecules play a key role in cell contraction, cell division, intracellular transport, and material transport along the axons of nerve cells, \emph{etc}~\cite{Alberts2007}. 

\begin{figure}[htbp]
  \centering
  \includegraphics[width=0.8\linewidth]{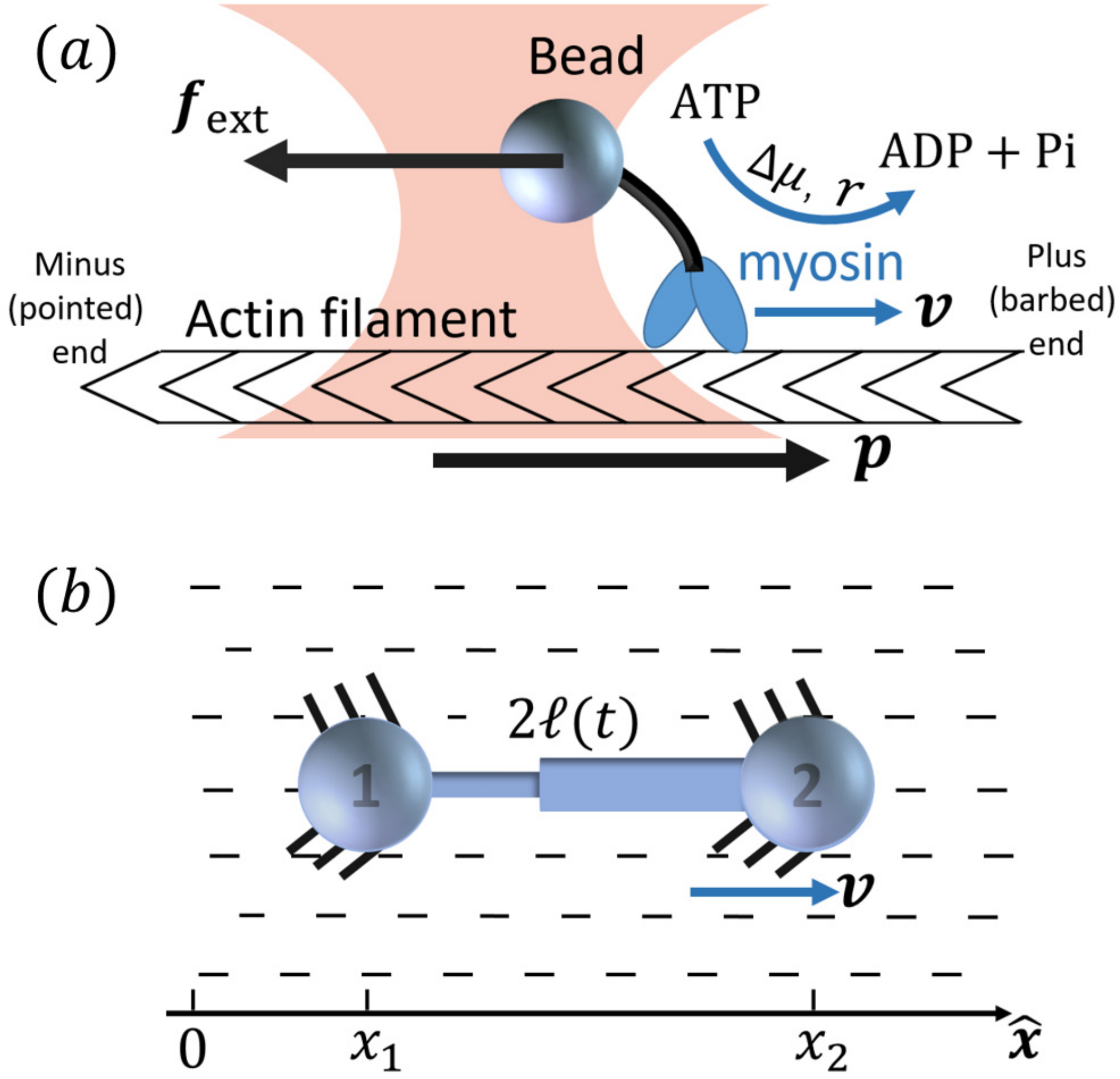} 
  \caption{(Color online) Directional motion of individual active units. (a) A myosin motor catalyzes ATP hydrolysis and converts the released chemical energy into its directional motion on an actin filament~\cite{Phillips2012} toward the plus (barbed) end. Meanwhile, an external force, $\bm{f}_{\mathrm{ext}}$, is applied on the myosin through optical trapping of a nano-bead attached to the motor. (b) A toy two-sphere microswimmer swims in viscous fluids. The body length, $2\ell(t)$, of the microswimmer oscillates cyclically and the friction coefficient, $\zeta$, is asymmetric for forward motion (with smaller friction) and backward motion (with larger friction), as shown in Eq.~(\ref{eq:Appl-ActiveUnit2-zeta}). The front-back asymmetry in friction is indicated by inclined thorns on the microspheres. }
\label{Fig:ActiveUnits}   
\end{figure}

In this subsection, we use OVP to formulate a thermodynamic description in the linear (near-equilibrium) regime for the directional motion of a translationary molecular motor along a polar filament against an external force, as shown in Fig.~\ref{Fig:ActiveUnits}a. This Onsager-type description is pioneered by Kedem \& Caplan~\cite{Kedem1965} and extended by Chen \& Hill~\cite{Hill1974}. We take our thermodynamic system to include the molecular motor and the surrounding solution of ATP, ADP, and $\mathrm{P_i}$. The system is coupled to a heat reservoir and a work reservoir, which can apply external forces on the motor, for example, by optical tweezers, that is, optical trapping of a nano-probe attached to the motor (see Fig.~\ref{Fig:ActiveUnits}a). The states of the thermodynamic system can be described by the average motor position $\bm{x}$, the polarization vector $\bm{p}$ (describing the polarity of the filament and assuming to point from minus end to plus end), and the average number $N_{\alpha}$ of chemical components involved in ATP hydrolysis with $\alpha={\mathrm{ATP}}$, ADP, and $\mathrm{P}_i$.

The reaction free energy for ATP hydrolysis takes the form of
\begin{equation}\label{eq:Appl-ActiveUnit3-Fr} 
{\cal F}_r={\cal F}_r(N_{\mathrm{ATP}},\, N_{\mathrm{ADP}}, \, N_{\mathrm{P_i}}),
\end{equation} 
from which we find the rate of the change of free energy as~\cite{Prost2015}
\begin{equation}\label{Eq:Appl-ActiveUnit3-Fdot}
\dot{\cal F}_r=-r\Delta\mu.
\end{equation} 
Here $r\equiv d\xi/dt$ is the reaction rate of ATP hydrolysis with $\xi$ being the reaction extent, $d\xi\equiv dN_{\alpha}/\nu_{\alpha}=-dN_{\mathrm{ATP}}=dN_{\mathrm{ADP}}=dN_{\mathrm{P_i}}$, and $\nu_{\alpha}$ being the stoichiometric coefficients (negative for reactants and positive for products, here $\nu_{\mathrm{ATP}}=-1$, $\nu_{\mathrm{ADP}}=+1$, and $\nu_{\mathrm{P_i}}=+1$). 
$\Delta \mu$ is the reaction affinity of ATP hydrolysis, given by
\begin{equation}\label{eq:Appl-ActivePolar-Deltarmu} 
\Delta \mu\equiv -\sum_{\alpha}\mu_{\alpha}\nu_{\alpha}=\mu_{\mathrm{ATP}}-\mu_{\mathrm{ADP}}-\mu_{\mathrm{P_i}},
\end{equation}
which measures the free-energy change for the hydrolysis of each ATP molecule. At chemical equilibrium $\Delta \mu=0$, whereas it is positive when ATP is in excess and negative when ADP is in excess. Under \emph{in vivo} conditions, ATP is usually in excess with $\Delta \mu>0$; the reason may be that ATP has evolved as a biological hydrotrope to keep biomolecules soluble at high concentrations and subsequently used a the ``energy currency'' of the cell due to its high energy phospho-diester bonds~\cite{Hyman2017}. 

The irreversible dynamics of the thermodynamic motor/filament system is characterized by two rates: the reaction rate $r$ and the average motor velocity $\bm{v}=\dot{\bm{x}}$. In the linear response regime close to equilibrium, the dissipation function is a quadratic function of the rates given by
\begin{equation}\label{Eq:Appl-ActiveUnit3-Phi}
\Phi= \frac{1}{2}\Lambda r^2-\lambda r\bm{p}\cdot \bm{v}+\frac{1}{2}\zeta \bm{v}^2,
\end{equation} 
\textcolor{blue}
and the rate of work done by the external force $\bm{f}_{\mathrm{ext}}$ to the motor is given by
\begin{equation}\label{Eq:Appl-ActiveUnit3-Wa}
\dot{W}_{\mathrm{ext}}= \bm{f}_{\mathrm{ext}}\cdot \bm{v}.
\end{equation} 
Here $\Lambda$ and $\zeta$ are generalized friction coefficients, and $\lambda$ is the mechanochemical coupling coefficient, which is nonzero only if the filaments are polar and can be either positive or negative~\cite{Julicher1997}. The positive-definiteness of the dissipation function requires $\Lambda,\, \zeta>0$ and $\Lambda \zeta-\lambda^2>0$. The degree of mechanochemical cross-coupling can be quantified by $q\equiv \lambda/\sqrt{\Lambda\zeta}$ (with $-1< q< 1$), as suggested by Kedem and Caplan~\cite{Kedem1965}. 

Using the Rayleighian ${\cal R}(r,\bm{v})=\dot{\cal F}_r+\Phi-\dot{W}_{\mathrm{ext}}$ in the presence of the external force
and Eqs.~(\ref{Eq:Appl-ActiveUnit3-Fdot})--(\ref{Eq:Appl-ActiveUnit3-Wa}), we minimize ${\cal R}$ with respect to $r$ and $\bm{v}$ and obtain 
\begin{subequations}\label{Eq:Appl-ActiveUnit3-Eqs}
\begin{equation}\label{Eq:Appl-ActiveUnit3-Eq1}
\Delta \mu=\Lambda r-\lambda v,
\end{equation} 
\begin{equation}\label{Eq:Appl-ActiveUnit3-Eq2}
{f}_{\mathrm{ext}}=-\lambda r+\zeta v.
\end{equation} 
\end{subequations} 
Here we have taken $\bm{v}$ and $\bm{f}_{\mathrm{ext}}$ to be in parallel with $\bm{p}$; a positive $v$ (or ${f}_{\mathrm{ext}}$) means the motion of the motor (or the direction of the external force) is along the direction of $\bm{p}$, pointing toward plus end as assumed. Note that the Onsager reciprocal relation for mechanochemical cross-coupling is automatically satisfied. 
Physically, Eq.~(\ref{Eq:Appl-ActiveUnit3-Eq2}) can be written as a balance equation ${f}_{\mathrm{ext}}+f_a+f_{\mathrm{fric}}=0$
for the external force ${f}_{\mathrm{ext}}$, 
the active force $f_a=\lambda r$, 
and the frictional force $f_{\mathrm{fric}}=-\zeta v$. 

To be specific, for the \emph{in vivo} motion of myosin motors  along actin filaments, ATP is usually in excess with a constant $\Delta \mu >0$ and myosin motor always move towards to the plus end of the actin filament. Therefore, the load-free motor velocity $v_0=\frac{q^2}{\lambda(1-q^2)}\Delta \mu$ 
(for $f_{\mathrm{ext}}=0$) 
must be positive (and hence $\lambda>0$ and $0<q<1$). In this case, we can identify the following four regimes from Eqs.~(\ref{Eq:Appl-ActiveUnit3-Eqs}):

(i) For $f_{\mathrm{ext}}>0$, we have $v>0$ and $r>0$ (hence $f_{\mathrm{ext}}v>0$ and $r\Delta\mu>0$), the external force pulls the motor to the plus end. 
Meanwhile, the excess ATP hydrolyzes and the chemical energy is consumed to drive the motor along the same direction to the plus end. 

(ii) For small negative force $f_{\mathrm{ext}}<0$ and $f_{\mathrm{ext}}>f_{\mathrm{stall,v}}$, we still have $v>0$ and $r>0$ (hence $f_{\mathrm{ext}}v<0$ and $r\Delta\mu>0$), the excess ATP hydrolyzes and the released chemical energy is converted into mechanical work. Here $f_{\mathrm{stall,v}}=-\lambda \Delta \mu/\Lambda<0$ is called stall force of myosin motion and when $f_{\mathrm{ext}}=f_{\mathrm{stall,v}}$, the motor is stationary with $v=0$. 

(iii) For $f_{\mathrm{stall,r}}<f_{\mathrm{ext}}<f_{\mathrm{stall,v}}$, the moving direction of the motor is reversed with $v<0$ and hence $f_{\mathrm{ext}}v>0$. That is, the external force is doing positive work on the motor moving along the plus end. However, we still have $r>0$ and hence $r\Delta\mu>0$. That is, the excess ATP hydrolyzes and the released chemical energy also drives the motion the motor towards the same plus end. Here $f_{\mathrm{stall,r}}=-\Delta\mu\zeta/\lambda<0$ is called stall force of ATP hydrolysis and when $f_{\mathrm{ext}}=f_{\mathrm{stall,r}}$, the ATP hydrolysis is inhibited with $r=0$. 
 
(iv) For $f_{\mathrm{ext}}<f_{\mathrm{stall,r}}<0$, we have $v<0$ and $r<0$, hence $f_{\mathrm{ext}}v>0$ and $r\Delta\mu<0$. That is, the external force is doing positive work on the motor and produces ATP that is already in excess; the system then works as an ATP pump.

Therefore, the motor/filament system is a reversible machine: it can not only convert chemical energy into mechanical work, but can also convert mechanical work into chemical energy. In this work, we are particularly interested in the regime (ii), in which ATP hydrolysis occurs spontaneously and the released chemical energy is used to drive the system out of equilibrium continuously. 


Now let's consider a practical limit at which the ATP hydrolysis rate $r$ (or equivalently the active force $f_a=\lambda r$) 
is taken as a given positive parameter that measures the activity of the system. 
That is, the effect of mechanical forces on the ATP hydrolysis is neglected and 
the rate of ATP hydrolysis is determined dominantly 
by the chemical affinity as $r\approx \Lambda^{-1}\Delta \mu>0$. 
This leads to a reduced description in which the position of the molecular motor in directional motion 
becomes the only state variable, 
while the amounts of the reactants and products in the ATP hydrolysis are no longer involved. 
As a result, the Rayleighian reduces to its extended form in Eq.~(\ref{eq:OVP-RayExt}) as
\begin{equation}\label{eq:Appl-ActiveUnit1-RayExt}
{\cal R}=\frac{1}{2}\zeta v^2 - f_a v - {f}_{\mathrm{ext}} v,    
\end{equation}
in which $\frac{1}{2}\zeta v^2 = \Phi$ is the dissipation function, 
$f_a v = \dot{W}_a$ is the rate of work done by the active force to the reduced system, and 
${f}_{\mathrm{ext}} v = \dot{W}_{\mathrm{ext}}$ 
is the rate of work done by the external force. 
Minimizing this Rayleighian gives the force balance equation $-\zeta v + f_a + f_{\mathrm{ext}}=0$. 

Finally, we would like to point out that the results obtained here from thermodynamic description for the motion of molecular motors on polar filaments are completely independent of any underlying microscopic mechanisms. However, the above linear-response theory applies only to the linear regime near equilibrium where $\Delta \mu/k_BT\ll 1$ and $f_{\mathrm{ext}}\xi/k_BT\ll 1$ with $\xi$ being the typical molecular size of relevant proteins. In real life, molecular motors mostly operate far from equilibrium (with $\Delta \mu\sim 10 k_{B} T$) and the velocity $v\left(f_{\mathrm{ext }}, \Delta \mu\right)$ and the rate of ATP consumption $r\left(f_{\mathrm{ext }}, \Delta \mu\right)$ are in general highly nonlinear. Therefore, more specific models such as a minimal two-states model for molecular motors should be constructed to arrive at a more comprehensive understanding of the specificity and robustness of the directional motion of motors in highly fluctuating environment~\cite{Julicher1997}.  
 

\subsection{A toy two-sphere microswimmer: active shape changes}\label{Sec:Appl-ActiveUnit2}

Many animals and cells can actively change their shape in some periodic or cyclic manner to migrate on frictional substrates or swim in their surrounding viscous fluid environment~\cite{Astley2015,Lauga2009,Elgeti2015,Ewa2020,Liverpool2020}. For example, snakes and some worms migrate on the ground by generating body waves to change their shapes~\cite{Astley2015}. Some bacteria such as Escherichia Coli, swims in fluids through bundling and rotating their flagella as driven by rotary motors~\cite{Lauga2009,Elgeti2015}. Many types of animal cells can also migrate on substrates by dramatic periodical shape changes~\cite{Ewa2020} as primarily driven by their active cellular cortex that consumes chemical energy. 
\textcolor{black}{It is interesting to note that in most of the cell migration driven by active shape changes, the cell migration velocity shows highly nonlinear dependence on the active force or the active shape-changing velocity of the cell. This seems to go far beyond the linear-response regime and be out of the scope of OVP. However, we will show that OVP developed in the linear-response regime can still be employed if we expand the set of state variables~\cite{Doi2020} to include not only the center-of-mass position of the cell but also the fast changing body length which describes the shape change.}

Specifically, to show how periodic shape changes can generate directional self-propulsion, here we consider a toy microswimmer that is composed of two microspheres~\cite{Zimmermann2009} as shown in Fig.~\ref{Fig:ActiveUnits}b. Let $x_1$ and $x_2$ denote the coordinates of these two microspheres. Then the directional motion and the shape changes of the microswimmer can be described by the temporal evolution of the center-of-mass position $x_c(t)=\frac{1}{2}(x_1+x_2)$ and the half-body-length of the swimmer  $\ell(t)=\frac{1}{2}(x_2-x_1)$, \textcolor{black}{which are taken as the two slow variables}. 
The toy microswimmer can actively change its shape by periodically changing its body length (or the distance between the two spheres) $2\ell(t)$ as  
\begin{equation}\label{eq:Appl-ActiveUnit2-ellt} 
\ell(t)=\ell_0+a\sin(\omega t),  
\end{equation}
where $\ell_0$ is the half of the average body length of the microswimmer, $a$ and $\omega$ are the amplitude and frequency of the shape oscillations, respectively, and the shape-oscillation period is $T_{\ell}=2\pi/\omega$. 

The toy microswimmer subjected to the periodic shape oscillations can achieve directional motion only when there exists some mechanisms that break the front-back symmetry. Here we consider an asymmetry in the viscous friction,  defined by $f_v=-\zeta(\dot{x}) \dot{x}$, in which the friction coefficient $\zeta(\dot{x}) $ depends on the moving direction of each microsphere according to:
\begin{align}\label{eq:Appl-ActiveUnit2-zeta}
\zeta(\dot{x}) = 
\begin{cases} 
\zeta_+,  &\mbox{if} \quad \dot{x} >0 \\
\zeta_-, & \mbox{if} \quad \dot{x}<0
\end{cases}.
\end{align}
The dynamics of the microswimmer is characterized by the velocities (the rates), $\dot{x}_1$ and $\dot{x}_2$, of the two spheres. To the leading order in the two rates, the dissipation function is given by
\begin{equation}\label{eq:Appl-ActiveUnit2-Phit} 
\Phi_t=\frac{1}{2}\zeta_1 \dot{x}_1^2+\frac{1}{2}\zeta_2\dot{x}_2^2
=\frac{1}{2}\zeta_1 (v-\dot{\ell})^2+\frac{1}{2}\zeta_2(v+\dot{\ell})^2,
\end{equation}
where $v=\dot{x}_c$ is the center-of-mass velocity of the microswimmer, $\zeta_1$ and $\zeta_2$ are the frictional coefficients given in Eq.~(\ref{eq:Appl-ActiveUnit2-zeta}) that depend on the signs of $\dot{x}_1=v-\dot{\ell}$ and $\dot{x}_2=v+\dot{\ell}$, respectively. Note that the dissipation function $\Phi_t$ is actually not quadratic in dissipative rates any more but highly nonlinear; a highly nonlinear dissipation function has been discussed in classical mechanics of particles before~\cite{Lurie2002}.

In most microswimmers, it is natural to assume that there is a clear separation of time scales between their shape oscillations and the directional motion. The directional motion of the microswimmer is usually much slower than its shape oscillations, \emph{i.e.}, $\ell_0/v\gg T_{\ell}$. We can, therefore, integrate out the relatively fast varying variable, the half-body-length $\ell(t)$, in one cycle of shape oscillation and arrive at a time-averaged dissipation function of the slow variable, the center-of-mass velocity $v$, by
$\Phi=\frac{1}{T_{\ell}}\int_0^{T_{\ell}} \Phi_t dt$ as 
\begin{equation}\label{eq:Appl-ActiveUnit2-Phi} 
\Phi(v)=\zeta_-\left[(1-{\cal A}\theta_s)(v^2+\frac{1}{2}a^2\omega^2)-\frac{3{\cal A}}{2\pi}v\sqrt{a^2\omega^2-v^2}\right],
\end{equation}
where $\theta_s=1-\frac{1}{\pi}\arccos(v/a\omega)$ for $|v|\le a\omega$ and ${\cal A}=1-\zeta_+/\zeta_-$ is a dimensionless parameter measuring the degree of the front-back asymmetry in the friction coefficients. 
\textcolor{black}{Note that this dissipation function is not quadratic but highly nonlinear in the rate $v$. 
We can define the Rayleighian ${\cal R}(v)=\Phi-\Phi(v=0)$ to determine the time-averaged directional motion of 
the toy microswimmer. Minimization of ${\cal R}(v)$ gives the cell migration velocity $v$ 
as a function of the active velocity $a\omega$ of the periodic shape change, and 
this function can be compared to experimental observations.} 
Furthermore, in the limit of small ${\cal A}$, we have $|v|\ll a\omega$ and the Rayleighian can be approximated, to the leading order, by
\begin{equation}\label{eq:Appl-ActiveUnit2-PhiLin} 
{\cal R}(v)\approx \frac{1}{2}\zeta_{\mathrm{eff}}v^2-f_{a,\mathrm{eff}}v,
\end{equation}
where $\zeta_{\mathrm{eff}}=\zeta_-+\zeta_+$ is the effective frictional coefficient and $f_{a,\mathrm{eff}}=\frac{2 a \omega}{\pi} (\zeta_--\zeta_+)$ is a time-averaged effective active force that drives the directional self-propulsion of the microswimmer. Minimization of ${\cal R}$ gives the directional velocity of the microswimmer as
\begin{equation}\label{eq:Appl-ActiveUnit2-vs} 
v =f_{a,\mathrm{eff}}/\zeta_{\mathrm{eff}},
\end{equation}
which approaches $v = a \omega{\cal A}/{\pi}$ in the limit of weak asymmetry with $\zeta_+/\zeta_-\to 1^-$ or ${\cal A} \to 0^+$.

We would like to give some remarks on the directional motion of the toy two-sphere microswimmer as follows.

\textcolor{black}{(i) Most dynamic behaviors of biological systems show strong nonlinearity. 
For example, in the toy two-sphere microswimmer, the swimmer migration velocity shows highly nonlinear dependence on the active force or the active shape-changing velocity of the swimmer. However, in many cases, OVP can still be employed if we expand the set of slow state variables  properly~\cite{Doi2020}. For example, here our set of slow variables includes not only the center-of-mass position of the swimmer 
but also its fast-changing body length.}

(ii) The dissipation function is non-zero even for symmetric microswimmers (with $\zeta=\zeta_-=\zeta_+$) when there is no average directional motion (\emph{i.e.}, $v=0$): $\Phi_0=\Phi(v=0)=\frac{1}{2}\zeta a^2\omega^2$. It arises in the symmetric microswimmer from the viscous dissipation due to the fast shape-oscillation in viscous fluids. 

(iii) Similar to the walk of molecular motors in the previous example, the active shape changes of microswimmers are also driven by spontaneous ATP hydrolysis. Then the irreversible dynamics of the microswimmer should be characterized by the rate of ATP hydrolysis $r$ in addition to the sphere velocities, $\dot{x}_1$ and $\dot{x}_2$. The rate of the change of free energy is given in Eq.~(\ref{Eq:Appl-ActiveUnit3-Fdot}) by $\dot{\cal F}_r=-r\Delta\mu$. The dissipation function is given, to the leading order in the rates, by
\begin{equation}\label{eq:Appl-ActiveUnit2-Phitr1} 
\Phi_t=\frac{1}{2}\Lambda r^2-\lambda r\dot{x}_1+\lambda r\dot{x}_2+\frac{1}{2}\zeta_1 \dot{x}_1^2+\frac{1}{2}\zeta_2\dot{x}_2^2.
\end{equation}
Also as mentioned in the previous example, in many practical cases we can take a  limit at which the ATP hydrolysis rate $r$ and hence the active force $f_a=\lambda r$ is a given positive parameter.  
This leads to a reduced description in which the dynamics of the microswimmer is described only by the sphere velocities. In this case, the Rayleighian takes to the general form of Eq.~(\ref{eq:OVP-Ray}) as
\begin{equation}
{\cal R}=\frac{1}{2}\zeta_1 \dot{x}_1^2+\frac{1}{2}\zeta_2\dot{x}_2^2 -f_a\dot{x}_1+f_a\dot{x}_2,
\end{equation}
in which the first two terms compose the dissipation function $\Phi(\dot{x}_1,\dot{x}_2)$, the last two terms compose the rate of work done by the active forces to the system with $\dot{W}_a=f_a\dot{x}_1+(-f_a)\dot{x}_2$. Minimizing this Rayleighian with respect to $\dot{x}_1$ and $\dot{x}_2$ gives the force balance equations for each sphere:
\begin{equation}\label{eq:Appl-ActiveUnit2-x1x2} 
f_a -\zeta_1 \dot{x}_1=0, \quad -f_a-\zeta_2\dot{x}_2=0,
\end{equation}
respectively. Here the pair of active forces $f_a$ and $-f_a$ forms a force dipole on the microswimmer. If the rate of the ATP hydrolysis $r$ is oscillating and the resulted active force $f_a=\lambda r$ takes the form of
\begin{equation}\label{eq:Appl-ActiveUnit2-fa} 
f_a=-\frac{2\zeta_-\zeta_+}{\zeta_-+\zeta_+}a\omega \cos(\omega t),
\end{equation}
then it will drive a shape oscillation defined in Eq.~(\ref{eq:Appl-ActiveUnit2-ellt}). Note that the effective active force $f_{a,\mathrm{eff}}$ in Eq.~(\ref{eq:Appl-ActiveUnit2-PhiLin}) is not a simple time-average of the oscillating active force $f_a$ in Eq.~(\ref{eq:Appl-ActiveUnit2-fa}), but is the ``net'' active force that drives the directional motion of the microswimmer.

Finally, we would like to point out that in our toy two-sphere microswimmer, the two necessary conditions for a steady directional motion are the active shape oscillations as the energy input and the frictional asymmetry that breaks the front-back symmetry. Similarly, for a long thin swimming micro-filament, the hydrodynamic friction is anisotropic: it experiences less friction when moving along its axis than perpendicular to it. In this case, a cyclic beat pattern on the filament will be able to drive directional propulsion in a similar manner as in the above one-dimensional toy microswimmer\footnote[2]{Private communications with M. Doi.}. 



\section{Applications 2: Active polar fluid models for collectives of active units}
\label{Sec:Appl-ActivePolar}

\begin{figure}[htbp]
  \centering
  \includegraphics[width=0.8\linewidth]{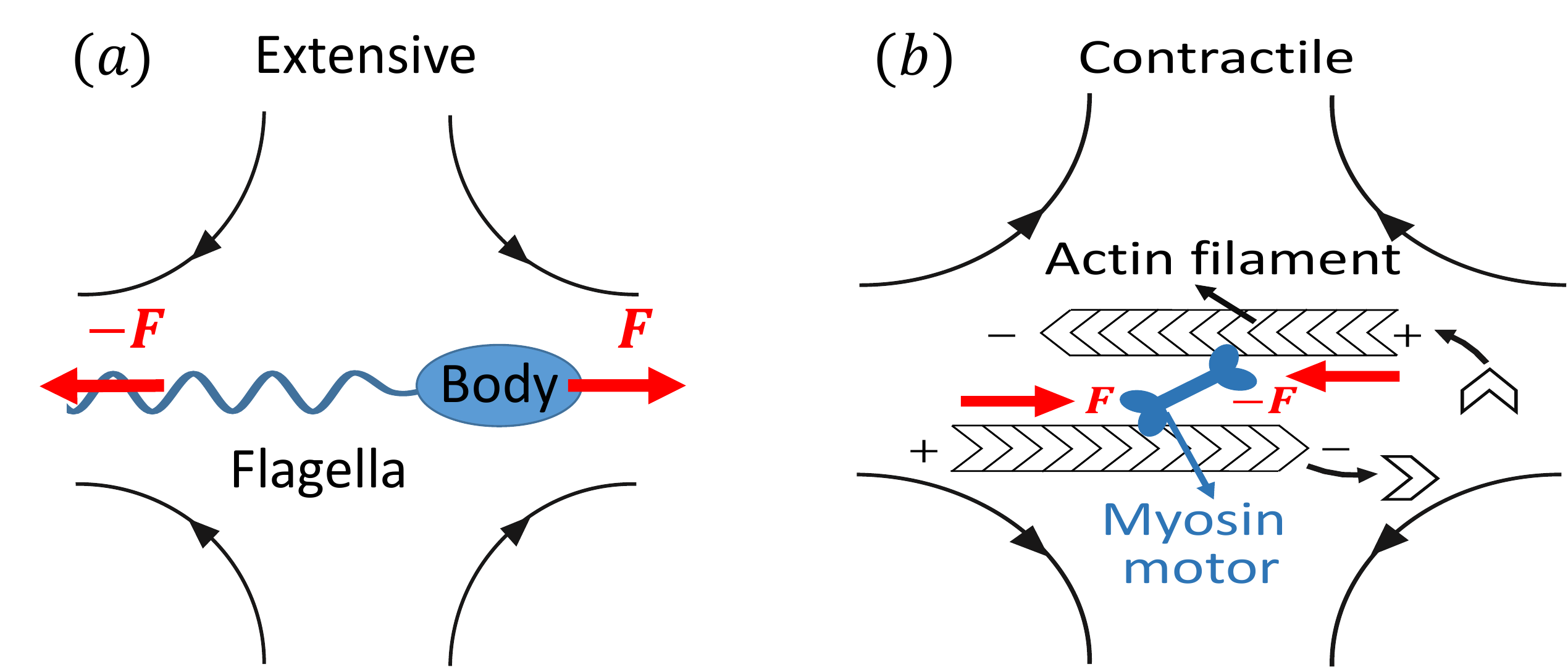} 
  \caption{(Color online) Active stresses generated by active units: (a) extensile force dipoles generated by the bacterial microswimmer, and (b) contractile force dipoles generated by the actomyosin filament.}
\label{Fig:forcedipoles}   
\end{figure}

Let's now consider an active fluid that includes collectives of active units with anisotropic shapes and polarity such as rod-like self-propelled colloids~\cite{Bechinger2016}, rod-like bacteria~\cite{Bechinger2016,Lauga2009,Elgeti2015}, and the active networks of stiff filaments in the cytoskeleton of living cells~\cite{Marchetti2013,Sam2013a}. The constituting active units can assume chemical energy to apply (extensile or contractile) dipole forces to their surrounding inert environment that drive the system locally out of equilibrium, as schematically shown in Fig.~\ref{Fig:forcedipoles}. Such active fluids are in contrast to more familiar passive (inert) non-equilibrium systems which are usually driven externally at their boundaries~\cite{Hohenberg1993}. 

\subsection{Active polar fluid regarded as a reactive fluid}
\label{Sec:Appl-ActivePolar-Reactive}

Active fluids with polar constituent agents often show phase separation (with coexisting dilute and dense phases) and collective orientational or polar order (with collective alignment on average)~\cite{Marchetti2013,Cates2018}. Such an active polar fluid can be viewed as a \emph{reactive} fluid where viscous flows and diffusion are closely coupled with biochemical reactions~\cite{Marchetti2013,Prost2015}. A generalized hydrodynamic theory has been developed to describe the dynamics of such active polar fluids, phenomenologically based on conservation laws and symmetry considerations~\cite{Prost2015,Marchetti2013}, which is in complete compatibility with micro-reversibility and Onsager's reciprocal relations. This is in contrast to other methods of modeling active polar fluids where some active (non-equilibrium) terms are selectively added to the dynamical equations for their passive counterparts to break TRS~\cite{Cates2014,Cates2015b,Cates2018,Thiele2020}.  
\textcolor{black}{In these methods, the rates of biochemical reactions in the active fluids are implicitly assumed to be a constant that is 
independent of the surrounding mechanical environment and simply determined by some preset reaction affinity. 
In this case, the active terms arising from biochemical reactions 
can not be derived from any free energy or dissipation function, and hence can only be added in an \emph{ad hoc} manner.}

\textcolor{black}{In this subsection, we present an alternative derivation of the generalized hydrodynamic model 
for an active polar fluid that is regarded as a reactive fluid involving the ATP hydrolysis/synthesis. 
In the next subsection, we will show that in the same active polar fluid, if the effects of polarization and flow on the ATP hydrolysis are negligible, then the rate of ATP hydrolysis becomes a constant 
simply determined by the preset constant chemical affinity $\Delta \mu$. 
The activities, driven by the spontaneous ATP hydrolysis, are then represented by local external non-conservative force fields 
that are added as the active terms to the dynamic model of a passive polar fluid.} 

\begin{figure}[htbp]
  \centering
  \includegraphics[width=0.8\linewidth]{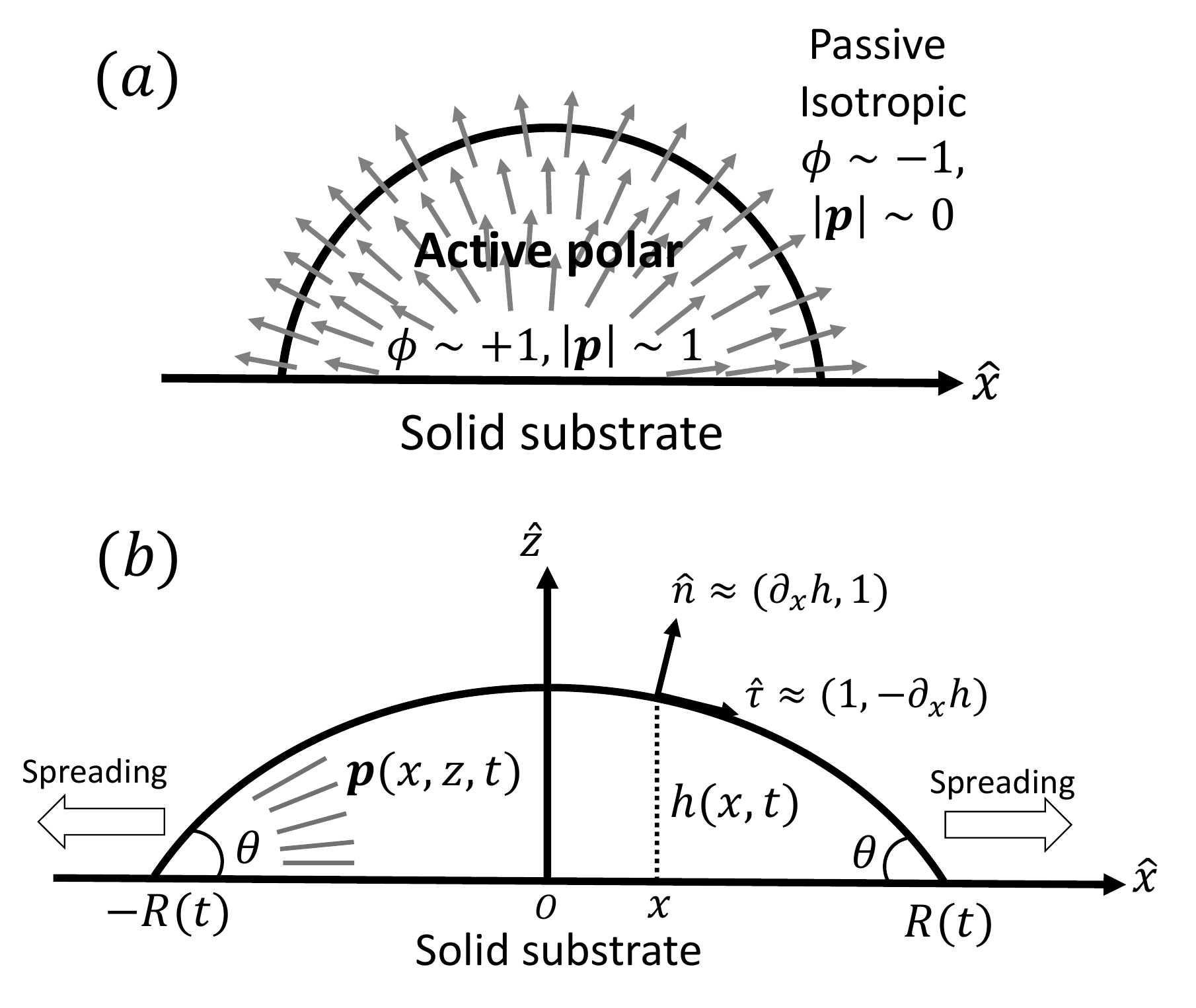} 
  \caption{A droplet containing a large number of active units on a solid substrate. (a) A general two-phase diffuse-interface model of an active droplet moving on solid substrates. A droplet of active polar phase is stabilized to coexist with surrounding fluids of passive isotropic phase. (b) Spreading of a thin active droplet on a solid substrate. Planar or homogeneous anchoring condition is assumed at all bounding surfaces with which the active units are in contact. }
\label{Fig:ActiveDrop}
\end{figure}

To be specific, here we use OVP to derive a diffuse-interface model for a droplet of active polar fluids moving on a solid substrate, as schematically shown in Fig.~\ref{Fig:ActiveDrop}a. The states of such an active polar droplet can be described by the following slow field variables: the scalar composition field $\phi({\bf r},t)$ (distinguishing the coexisting passive isotropic phase from the active polar phase), the polarization vector field $\bm{p}({\bf r},t)$ (describing the average orientation of active polar agents), the average fluid velocity field $\bm{v}({\bf r},t)$, and the density field $n_{\alpha}({\bf r},t)$ of chemical components involved in ATP hydrolysis (Eq.~(\ref{eq:Appl-ActiveUnit-ATP})) with $\alpha={\mathrm{ATP}}$, ADP, and $\mathrm{P}_i$. For an active polar fluid that is confined between solid substrates or flows at the solid surfaces, the total free energy includes four contributions, $\mathcal{F}\left[\phi, \bm{p}, n_{\alpha}\right]=\mathcal{F}_{\phi}+\mathcal{F}_{p}+\mathcal{F}_{r}+\mathcal{F}_{s}$, as respectively given by
\begin{subequations}\label{eq:Appl-ActivePolar-F} 
\begin{equation}\label{eq:Appl-ActivePolar-Fphi} 
\mathcal{F}_{\phi}\left[\phi\right]=\int d \bm{r}\left[-\frac{1}{2}a\phi^2+\frac{1}{4}a\phi^4+\frac{1}{2} K_{\phi}(\bm{\nabla}\phi)^{2}\right],
\end{equation}
\begin{equation}\label{eq:Appl-ActivePolar-Fp} 
\mathcal{F}_{p}\left[\phi,\,\bm{p}\right]=\int d \bm{r}\left[-\frac{1}{2} b_1(\phi) \bm{p}^{2}+\frac{1}{4} b_2 \bm{p}^{4}+\frac{1}{2} K_{p}(\bm{\nabla} \bm{p})^{2}+c \bm{p} \cdot  \bm{\nabla} \phi \right],
\end{equation}
\begin{equation}\label{eq:Appl-ActivePolar-Fr} 
\mathcal{F}_{r}\left[n_{\alpha}\right]=\int d \bm{r}f_r(n_{\mathrm{ATP}},\, n_{\mathrm{ADP}}, \, n_{\mathrm{P_i}}),
\end{equation}
\begin{equation}\label{eq:Appl-ActivePolar-Fs} 
\mathcal{F}_{s}\left[\phi,\,\bm{p}\right]=\int dA \left[a_s\phi +\frac{1}{2}b_s(\hat{\bm{p}}\cdot\hat{\bm{n}})^2\right].
\end{equation}
\end{subequations}
Here $\mathcal{F}_{\phi}$ is the free energy for the two (isotropic and polar) phase coexistence, in which $a$ and the stiffness parameter $K_{\phi}$ are both constants. 
$\mathcal{F}_{p}$ is the free energy for polar liquid crystallinity, in which for simplicity, we employ the approximation of one elastic constant~\cite{deGennes1993,Joanny2012} $K_{p}$ and we take $b_1(\phi)=-b_0\phi$ with $b_0>0$ that controls the isotropic-to-polar phase transition: $b_1<0$ in the polar phase (with $\phi=1$) and $b_1>0$ in the isotropic phase (with $\phi=-1$). Note that the last cross-coupling term in $\mathcal{F}_{p}$ represents the cases of perpendicular anchoring at the isotropic-polar phase interfaces and defines the orientation of $\bm{p}$: if $c>0$, $\bm{p}$ points from polar phase to isotropic phase. 
$\mathcal{F}_{r}$ is the reaction free energy for ATP hydrolysis: if $df_r/dn_{\mathrm{ATP}}>0$ (that is, ATP is in excess), the forward ATP hydrolysis is exergonic, occurs spontaneously, and can be used to drive the changes in the motor configurations and generate mechanical motion, resulting in active dipole forces on the the surrounding passive polar fluids (see Fig.~\ref{Fig:forcedipoles}). In this case, the polar fluid will never reach thermodynamic equilibrium states and will be driven locally out of equilibrium by the active units or the motors that consumes ATP. 
$\mathcal{F}_{s}$ is the surface energy at the substrate surfaces (with $a_s$ and $b_s$ being constants, and $A$ being the surface area), which characterizes the adhesion strength of active units to the surface and the anchoring conditions for the agent orientation. 
Note that such a free energy $\mathcal{F}$ will stabilize a droplet of active polar phase ($\phi \simeq 1$ and $|\bm{p}|\simeq 1$) in coexistence with surrounding fluids of passive isotropic phase ($\phi \simeq -1$ and $|\bm{p}|\simeq 0$), or \emph{vice versa}~\cite{Cates2018}.

The composition variable, $\phi$, is \textcolor{black}{a conserved phase parameter} and its dynamics follows the following conservation equation
\begin{equation}\label{eq:Appl-ActivePolar-phit} 
\partial_t \phi=-\nabla\cdot (\phi\bm{v}+\bm{J}).
\end{equation}
However, the polarization vector $\bm{p}$ is not conserved and its rate of change is defined by $\dot{\bm{p}} \equiv \partial \bm{p} / \partial t+\bm{v} \cdot \nabla \bm{p}$. Furthermore, the density fields $n_{\alpha}$ (with $\alpha={\mathrm{ATP}}$, ADP, and P) are also not conserved due to the presence of chemical reaction of ATP hydrolysis or synthesis, and $n_{\alpha}$ follows the dynamic equation of the form 
\begin{equation}\label{eq:Appl-ActivePolar-nt} 
\partial_t n_{\alpha}=-\nabla\cdot (n_{\alpha}\bm{v})+r\nu_{\alpha}.
\end{equation}
\textcolor{black}{Here $\nu_{\alpha}$ is the stoichiometric coefficients (see the discussion about ATP hydrolysis after Eq.~(\ref{Eq:Appl-ActiveUnit3-Fdot})). 
The reaction rate $r\equiv d\xi/dt$ depends on the concentrations of all chemical components. (This dependence is one of the constitutive relations to be derived later from OVP and 
given in Eq.~(\ref{eq:Appl-ActivePolar-Dyn1c}).) However, in the present work, we do not intend to go into the specific expression of this concentration dependence 
because our purpose is to show that accompanying the chemical reaction, a mechanical force arises from the mechanochemical coupling as the active force. 
It is also noted that the general dynamics of ATP hydrolysis should be described by reaction-diffusion equations. 
However, for simplicity, here we neglect the diffusion processes, assuming that 
the density of each component is simply advected by the flow and produced or consumed by the chemical reaction. }

Using Eqs.~(\ref{eq:Appl-ActivePolar-phit}) and~(\ref{eq:Appl-ActivePolar-nt}) and the definition of $\dot{\bm{p}}$, we obtain the change rate of free energy from Eqs.~(\ref{eq:Appl-ActivePolar-F}) as 
\begin{align}\label{eq:Appl-ActivePolar-Fdot} 
\dot{\mathcal{F}}=&\int d \bm{r}\left[-(\nabla\cdot \bm{\sigma}^{\mathrm{e}})\cdot\bm{v} +\nabla\mu_{\phi}\cdot\bm{J}-\bm{h}\cdot\dot{\bm{p}}-\Delta \mu r\right] \nonumber \\
&+\int dA \left[L\dot{\phi}+\bm{H}\cdot\dot{\bm{p}}-(L\nabla_{\tau}\phi+H_k\nabla_{\tau}p_k)\cdot \bm{v}_{\tau}\right]
\end{align}
where $\bm{\sigma}^{\mathrm{e}}$ is the Ericksen stress tensor~\cite{deGennes1993,Doi2011,Qian2008} given by
\begin{equation}\label{eq:Appl-ActivePolar-Sigmae} 
\bm{\sigma}^{\mathrm{e}}=-\hat{p} \bm{I}-K_{\phi}\nabla \phi\nabla\phi -K_{p} \nabla p_{k} \nabla p_{k}-c\bm{p}\nabla\phi,
\end{equation}
and $\bm{\sigma}^{\mathrm{e}}$ satisfies the generalized Gibbs-Duhem relation
\begin{equation}\label{eq:Appl-ActivePolar-GibbsDuhem} 
-\nabla \cdot \bm{\sigma}^{\mathrm{e}}=\phi \nabla \mu_{\phi} - \nabla \bm{h}\cdot \bm{p}+\sum_{\alpha}n_{\alpha}\nabla\mu_{\alpha}.
\end{equation}
Here $\hat{p}=-\hat{f}_{\phi}-\hat{f}_{p}-\hat{f}_r+\phi\mu_{\phi}+\bm{p}\cdot\bm{h}+\sum_{\alpha}n_{\alpha}\mu_{\alpha}$ is the generalized pressure with $\hat{f}_m$ ($m={\phi},\, {p},\, {r}$) being the volume density of free energy in Eqs.~(\ref{eq:Appl-ActivePolar-F}), $\mu_{\alpha}\equiv \partial f_r/\partial n_{\alpha}$ being the chemical potential, and the reaction affinity, $\Delta \mu$, of ATP hydrolysis given in Eq.~(\ref{eq:Appl-ActivePolar-Deltarmu}). The chemical potentials $\mu_{\phi}$ in the bulk and $L$ at the solid surfaces are, respectively, given by
\begin{equation}\label{eq:Appl-ActivePolar-muphi} 
\mu_{\phi} \equiv \frac{\delta \mathcal{F}}{\delta \phi}=-a(\phi-\phi^3)-K_{\phi}\nabla^2\phi-\frac{1}{2}b_1'\bm{p}^2-c\nabla\cdot \bm{p},
\end{equation}
\begin{equation}\label{eq:Appl-ActivePolar-Lsurf} 
L \equiv a_s+K_{\phi} \hat{\bm{n}} \cdot \nabla \phi+c\hat{\bm{n}} \cdot\bm{p},
\end{equation}
with $b_1'=d b_1/d\phi$ denoting the derivative of $b_1$. 
The molecular field $\bm{h}$ in the bulk and $\bm{H}$ at the solid surfaces are, respectively, given by 
\begin{equation}\label{eq:Appl-ActivePolar-h} 
\bm{h} \equiv -\frac{\delta \mathcal{F}}{\delta \bm{p}}=b_1 \bm{p}-b_2\bm{p}^{3}+K_{p} \nabla^{2} \bm{p}-c\nabla \phi, 
\end{equation}
\begin{equation}\label{eq:Appl-ActivePolar-Hsurf} 
\bm{H} \equiv b_s\bm{p}\cdot\hat{\bm{n}}\hat{\bm{n}}+K_{p} \hat{\bm{n}} \cdot \nabla \bm{p}.
\end{equation}

The energy dissipation function $\Phi$ is a quadratic function of three dissipative rates: the shear rate $\dot{\bm{\epsilon}}=\frac{1}{2}\left(\nabla \bm{v}+\nabla \bm{v}^T\right)$, the rate of change of polarization $\dot{\bm{p}}$, and the rate of ATP hydrolysis $r$. These rates have the same time parity and from symmetry considerations, $\Phi$ can be written into the following invariant scalar form as~\cite{Doi2011}
\begin{equation}\label{eq:Appl-ActivePolar-PhiF} 
\begin{aligned}
\Phi &=\int d\bm{r}
\left[\frac{1}{2}\beta_{1}(\dot{\bm{\epsilon}}:\bm{p}\bm{p})^2
+\frac{1}{2}\beta_{2}\dot{\bm{\epsilon}}^{2}
+\frac{1}{2}\beta_{3}(\dot{\bm{\epsilon}}\cdot\bm{p})^{2}
+\frac{1}{2}\beta_{4}\bm{P}^{2}\right.\\
& \left.+ \beta_{5}\dot{\bm{\epsilon}}:\bm{P}\bm{p}
+\frac{1}{2}\beta_{6}r^{2}
+ \beta_{7}r\bm{P}\cdot \bm{p} 
+ \beta_{8}r(\dot{\bm{\epsilon}}:\bm{p}\bm{p})
+\frac{1}{2}\beta_{9}\bm{J}^{2}\right]\\
& +\int dA\left[\frac{1}{2}\beta_{10}{\dot{\phi}}^{2}+\frac{1}{2}\beta_{11} (\bm{v}_{\tau}^{\mathrm slip})^{2}\right],
\end{aligned}
\end{equation}
where the frictional coefficients $\beta_1,\dots,\beta_{11}$ are constants and the resulted frictional coefficient matrix, $\bm{\beta}$, is positive definite. That is, the diagonal coefficient terms in $\bm{\beta}$ are all positive, but the off-diagonal coefficient terms that describe the cross-coupling effects between two dissipative processes can be \emph{negative} (although some other inequality relations have to be satisfied to keep the positive-definiteness of $\bm{\beta}$).
Note that the proper dissipative rates that are present in $\Phi$ and associated with the orientational dynamics of $\bm{p}$ is not $\dot{\bm{p}}$ but the convected co-rotational time derivative~\cite{deGennes1993,Doi2011,Qian2008,Doi2020} of $\bm{p}$:
\begin{equation}
\bm{P} \equiv \dot{\bm{p}} -\bm{\omega} \times \bm{p} =\dot{\bm{p}} +\bm{\Omega} \cdot \bm{p},   
\end{equation}
which characterizes the rotation of $\bm{p}$ relative to the rotation of surrounding fluids with $\omega_i=\frac{1}{2}\epsilon_{ijk}\Omega_{jk}$ being the vorticity, and $\Omega_{jk}=\frac{1}{2}\left(\partial_j v_k-\partial_k v_j\right)$ being the antisymmetric part of the velocity gradient tensor. Moreover, for simplicity, we have neglected the cross coupling among 
the transport of $\phi$, the dynamics of polarization, and the ATP hydrolysis in the bulk fluid. \textcolor{black}{This coupling can represent 
the treadmilling dynamics of the constituting components~\cite{Julicher2011}}. We have also ignored the possible dissipative relaxation processes associated with the anchoring of $\bm{p}$ at the solid surfaces. 
In addition, we would like to point out that the choice of dissipative rates in active polar fluids is not unique~\cite{deGroot1984,Gyarmati1970}. Another set of rates has been taken by Marchetti \emph{et al.}~\cite{Marchetti2013}: the viscous stress $\bm{\sigma}^v$ (or the momentum flux), $r$ and $\bm{P}$. In this case, the time parity of $\bm{\sigma}^v$ is different from the other two rates $r$ and $\bm{P}$. A brief discussion about the consequences of different choices of dissipative rates or thermodynamic fluxes on the symmetry of Onsager coupling matrix and the applications of OVP has been presented in the Appendix Sec.~\ref{Sec:AppA-TimeParity}. 

Then the Rayleighian is given by
\begin{equation}\label{eq:Appl-ActivePolar-Ray} 
{\cal R}[\bm{v},\bm{J},\dot{\bm{p}},r]=\dot{\cal F}[\bm{v},\bm{J},\dot{\bm{p}},r]+\Phi[\bm{v},\bm{J},\bm{P},r]-\int d\bm{r} p\nabla \cdot \bm{v},
\end{equation}
where the local incompressibility constraints, $\nabla \cdot \bm{v}=0$, have been taken into account with the pressure $p$ being the Lagrange multiplier. Minimization of ${\cal R}[\bm{v},\bm{J},\dot{\bm{p}},r]$ with respect to the rates gives the dynamic equations for two-phase active polar flows as
\begin{subequations}\label{eq:Appl-ActivePolar-Dyn1} 
\begin{equation}\label{eq:Appl-ActivePolar-Dyn1a} 
-\nabla p+\nabla\cdot\left(\bm{\sigma}^{\mathrm{e}}+\bm{\sigma}^{\mathrm{v}}+\bm{\sigma}^{\mathrm{a}}\right)=0,
\end{equation} 
\begin{equation}\label{eq:Appl-ActivePolar-Dyn1b}  
\bm{h}=\gamma_{1}\bm{P}+\gamma_2\bm{p}\cdot\dot{\bm{\epsilon}}-\bm{h}^{\mathrm{a}},
\end{equation}
\begin{equation}\label{eq:Appl-ActivePolar-Dyn1c}  
\Delta \mu=-\lambda\bm{p} \cdot \bm{P}-{\tilde{\zeta}}\bm{p}\bm{p}:\dot{\bm{\epsilon}}+\Lambda r, 
\end{equation}
\end{subequations}
together with the incompressibility condition $\nabla \cdot \bm{v}=0$, the dynamic equations for $\phi$ in Eq.~(\ref{eq:Appl-ActivePolar-phit}) and for $n_{\alpha}$ in Eq.~(\ref{eq:Appl-ActivePolar-nt}). 
The stress tensors, $\bm{\sigma}^{\mathrm{e}}$, $\bm{\sigma}^{\mathrm{v}}$, and $\bm{\sigma}^{\mathrm{a}}$, are, respectively, given by Eq.~(\ref{eq:Appl-ActivePolar-Sigmae}), and
\begin{subequations}\label{eq:Appl-ActivePolar-Sigmava} 
\begin{align}\label{eq:Appl-ActivePolar-Sigmav} 
\bm{\sigma}^{\mathrm{v}}=\alpha_1(\dot{\bm{\epsilon}}:\bm{p}\bm{p})\bm{p}\bm{p}+\alpha_2\bm{p}\bm{P}+\alpha_3\bm{P}\bm{p}+\alpha_4\dot{\bm{\epsilon}}+\alpha_5\bm{p}\bm{p}\cdot\dot{\bm{\epsilon}}+\alpha_6\dot{\bm{\epsilon}}\cdot\bm{p}\bm{p},
\end{align}
\begin{equation}\label{eq:Appl-ActivePolar-Sigmaa} 
\bm{\sigma}^{\mathrm{a}}=-\tilde{\zeta} {r} \bm{p}\bm{p},
\end{equation}
\end{subequations}
with $\alpha_i$ being the Leslie viscosity coefficients. Note that $\hat{p}$ in $\bm{\sigma}^{\mathrm{e}}$ in Eq.~(\ref{eq:Appl-ActivePolar-Sigmae}) can be absorbed into $p$ in Eq.~(\ref{eq:Appl-ActivePolar-Dyn1a}) due to the incompressibility of the active polar fluids.
The active molecular field $\bm{h}^{\mathrm{a}}$ is driven by ATP hydrolysis and is defined via 
\begin{equation}\label{eq:Appl-ActivePolar-ha} 
\bm{h}^{\mathrm{a}}\equiv \lambda r \bm{p}.    
\end{equation}
The diffusion flux in Eq.~(\ref{eq:Appl-ActivePolar-phit}) is given by $\bm{J}=-M\nabla \mu_{\phi}$. Note that the mechanochemical cross-coupling indicated in Eqs.~(\ref{eq:Appl-ActivePolar-Dyn1}) is similar (although more complex) to Eqs.~(\ref{Eq:Appl-ActiveUnit3-Eqs}) for the mechanochemical coupling in the motor/filament system in Sec.~\ref{Sec:Appl-ActiveUnit1}. The general discussions there about the reversible conversion of chemical energy and mechanical work also apply here in active polar fluids. We will also focus on the regime where ATP is in excess and its hydrolysis occurs spontaneously and the released chemical energy is used by the suspending active units to drive the surrounding inert polar fluids out of equilibrium continuously.

Furthermore, from the minimization of ${\cal R}[\bm{v},\bm{J},\dot{\bm{p}},r]$, we can also obtain the thermodynamically-consistent boundary conditions at the solid surfaces that supplements the dynamic equations in the bulk fluids: 
\begin{subequations}\label{eq:Appl-ActivePolar-BCs1} 
\begin{equation}\label{eq:Appl-ActivePolar-BCs1a} 
\partial_t \phi+ \bm{v}_{\tau}\cdot \nabla_{\tau} \phi= -\Gamma L(\phi,\bm{p}),
\end{equation}
\begin{equation}\label{eq:Appl-ActivePolar-BCs1b} 
\beta\bm{v}_{\tau}^{\mathrm slip}= -\hat{\bm{n}}\cdot \left(\bm{\sigma}^{\mathrm{v}}+\bm{\sigma}^{\mathrm{a}}\right)\cdot \hat{\bm{\tau}}+ L\partial_{\tau}\phi,
\end{equation} 
together with the impermeability boundary conditions,
\begin{equation}\label{eq:Appl-ActivePolar-BCs1c} 
\hat{\bm{n}}\cdot \bm{v}=0, \quad \hat{\bm{n}}\cdot \bm{J}=0,
\end{equation}
and the equilibrium anchoring boundary condition,
\begin{equation}\label{eq:Appl-ActivePolar-BCs1d} 
\bm{H} \equiv b_s\bm{p}\cdot\hat{\bm{n}}\hat{\bm{n}}+K_{p} \hat{\bm{n}} \cdot \nabla \bm{p}=0.
\end{equation}
\end{subequations}
\textcolor{black}{Here the subscript $\tau$ denotes the tangential component in the plane of the substrate. }
Note that the boundary condition for the tangential velocity in Eq.~(\ref{eq:Appl-ActivePolar-BCs1b}) 
is similar to the generalized Navier boundary condition (GNBC) formulated for immiscible two-phase flows at solid surfaces~\cite{Qian2006}. 
The frictional coefficient $\beta$ now depends on the adhesion strength of the active fluids on the substrate surfaces.

The phenomenological coefficients in the above equation systems are functions of the frictional coefficients $\beta_i$ in the dissipation function $\Phi$ in Eq.~(\ref{eq:Appl-ActivePolar-PhiF}) as~\cite{Doi2011}
\begin{align}
\centering
&\alpha_{1}=\beta_1, \quad \alpha_{2}=-\frac{1}{2}\beta_4+\frac{1}{2}\beta_{5}, \quad \alpha_{3}=\frac{1}{2}\beta_4+\frac{1}{2}\beta_{5}, \quad \alpha_{4}=\beta_{2}, \nonumber \\
&\alpha_{5}=\frac{1}{2}\beta_3-\frac{1}{2}\beta_{5}, \quad \alpha_{6}=\frac{1}{2}\beta_3+\frac{1}{2}\beta_{5}, \quad \Lambda=\beta_6,\quad \lambda=-\beta_7, \nonumber \\
&\tilde{\zeta}=-\beta_{8}, \quad M=\beta_9^{-1}, \quad \Gamma=\beta_{10}^{-1}, \quad \beta=\beta_{11}, 
\end{align}
with $\gamma_1=\alpha_3-\alpha_2$ and $\gamma_2=\alpha_6-\alpha_5$. Note that the Parodi relation $\alpha_{2}+\alpha_{3}=\alpha_{6}-\alpha_{5}$ and the symmetry of coefficients for cross-coupling effects are all automatically satisfied. 

\subsection{Active polar fluid regarded as a passive polar fluid under local external non-conservative fields}
\label{Sec:Appl-ActivePolar-Force}
 
In the last part of Sec.~\ref{Sec:Appl-ActiveUnit1}, we have considered a practical and useful limit at which the effect of mechanical forces on the ATP hydrolysis can be neglected and the rate of ATP hydrolysis is determined dominantly by the chemical affinity as a constant parameter. Here we consider the same limit of the above general dynamics of active polar fluids: the effects of polarization and flow on the ATP hydrolysis are negligibly small, and the rate $r$ in Eq.~(\ref{eq:Appl-ActivePolar-Dyn1c}) is constant and simply determined by the reaction affinity or the change of reaction free energy, \emph{i.e.}, $r\approx \Lambda^{-1}\Delta \mu$. In this case, the rate of change of the free energy and the dissipation function reduce to
\begin{align}\label{eq:Appl-ActivePolar-Fdot2} \tilde{\dot{\mathcal{F}}}=&\int d \bm{r}\left[-(\nabla\cdot \bm{\sigma}^{\mathrm{e}})\cdot\bm{v} +\nabla\mu_{\phi}\cdot\bm{J}-\bm{h}\cdot\dot{\bm{p}}\right] \nonumber \\
&+\int dA \left[L\dot{\phi}+\bm{H}\cdot\dot{\bm{p}}-(L\nabla_{\tau}\phi+H_k\nabla_{\tau}p_k)\cdot \bm{v}_{\tau}\right],
\end{align}
\begin{align}\label{eq:Appl-ActivePolar-PhiF2}
\tilde{\Phi} &=\int d\bm{r}
\left[\frac{1}{2}\beta_{1}(\dot{\bm{\epsilon}}:\bm{p}\bm{p})^2
+\frac{1}{2}\beta_{2}\dot{\bm{\epsilon}}^{2}
+\frac{1}{2}\beta_{3}(\dot{\bm{\epsilon}}\cdot\bm{p})^{2}
+\frac{1}{2}\beta_{4}\bm{P}^{2}\right. \nonumber\\
& \left.+\frac{1}{2}\beta_{5}\dot{\bm{\epsilon}}:\bm{P}\bm{p}  
+\frac{1}{2}\beta_{7}r\bm{P}\cdot \bm{p} 
+\frac{1}{2}\beta_{8}r(\dot{\bm{\epsilon}}:\bm{p}\bm{p})+\frac{1}{2}\beta_{9}\bm{J}^{2}\right] \nonumber \\
&+\int dA\left[\frac{1}{2}\beta_{10}{\dot{\phi}}^{2}+\frac{1}{2}\beta_{11} (\bm{v}_{\tau}^{\mathrm slip})^{2}\right].
\end{align}
It is interesting to note that the two mechanochemical cross-coupling terms (the two terms with coefficients $\beta_{7}$ and $\beta_{8}$) in $\tilde{\Phi}$,  parametrized by the constant reaction rate $r$, can be rewritten as 
${-\dot{\cal W}_a}$, 
\begin{equation}\label{eq:Appl-ActivePolar-Wa} 
-\dot{\cal W}_a\equiv \int d\bm{r} \left[ \frac{1}{2}\beta_{7}r\bm{P}\cdot \bm{p} 
+\frac{1}{2}\beta_{8}r(\dot{\bm{\epsilon}}:\bm{p}\bm{p})\right]=\int d\bm{r} \left[ \bm{\sigma}^{\mathrm{a}}:\nabla\bm{v} - \bm{h}^{\mathrm{a}}\cdot\dot{\bm{p}}\right],
\end{equation}
in which the active stress $\bm{\sigma}^{\mathrm{a}}$  defined in Eq.~(\ref{eq:Appl-ActivePolar-Sigmaa}) and the active molecular field $\bm{h}^{\mathrm{a}}$ defined in Eq.~(\ref{eq:Appl-ActivePolar-ha}) now become
\begin{subequations}\label{eq:Appl-ActivePolar-sigmaaha}
\begin{equation}\label{eq:Appl-ActivePolar-sigmaa}
\bm{\sigma}^{\mathrm{a}}=-\zeta\bm{p}\bm{p},   
\end{equation}
\begin{equation}\label{eq:Appl-ActivePolar-ha2}
\bm{h}^{\mathrm{a}}=\zeta_h\bm{p}, 
\end{equation}
\end{subequations}
with $\zeta\equiv \tilde{\zeta}\Lambda^{-1} \Delta \mu$ and $\zeta_h\equiv \lambda\Lambda^{-1} \Delta \mu$. The cross-coupling coefficients $\beta_7$ and $\beta_8$ (hence $\zeta$ and $\zeta_h$) can be either positive or negative as long as they can preserve the positive definiteness of the dissipation function. For example, negative and positive values of $\zeta$ correspond to contractile and extensile active stresses, respectively, as schematically shown in Fig.~\ref{Fig:forcedipoles}. 

We would like to further point out that Eq.~(\ref{eq:Appl-ActivePolar-Wa}) indicates that in the limit of constant reaction rate of ATP hydrolysis, the ATP-induced activity in the active polar fluid can be regarded simply as some local non-conservative fields  applied externally on the passive polar fluid. 
The active characteristic of these external fields is reflected in the fact that the active stress $\bm{\sigma}^{\mathrm{a}}$ and the active molecular field $\bm{h}^{\mathrm{a}}$ both depend on the local state variable (the polarization), $\bm{p}$. Furthermore, these active fields driven by spontaneous ATP hydrolysis break the time-reversal symmetry of the polar fluids. 

It follows that according to the general form of Eq.~(\ref{eq:OVP-Ray}), the Rayleighian in Eq.~(\ref{eq:Appl-ActivePolar-Ray}) can be rewritten as
\begin{equation}\label{eq:Appl-ActivePolar-Ray2} 
{\cal R}[\bm{v},\bm{J},\dot{\bm{p}}]={\tilde{\dot{\mathcal{F}}}}[\bm{v},\bm{J},\dot{\bm{p}}]+\hat{\Phi}[\bm{v},\bm{J},\bm{P}]-{\dot{\cal W}_a}-\int d\bm{r} p\nabla \cdot \bm{v}.
\end{equation}
Here $\dot{\cal W}_a$ represents the work power done on the passive polar fluids by the active stress $\bm{\sigma}^{\mathrm{a}}$ and the active molecular field $\bm{h}^{\mathrm{a}}$. The dissipation function $\hat{\Phi}$ for the passive polar fluid is given by 
\begin{align}\label{eq:Appl-ActivePolar-PhiF3}
\hat{\Phi} &=\int d\bm{r}
\left[\frac{1}{2}\beta_{1}(\dot{\bm{\epsilon}}:\bm{p}\bm{p})^2
+\frac{1}{2}\beta_{2}\dot{\bm{\epsilon}}^{2}
+\frac{1}{2}\beta_{3}(\dot{\bm{\epsilon}}\cdot\bm{p})^{2}
+\frac{1}{2}\beta_{4}\bm{P}^{2}\right. \nonumber \\
& \left.+\frac{1}{2}\beta_{5}\dot{\bm{\epsilon}}:\bm{P}\bm{p}  
+\frac{1}{2}\beta_{9}\bm{J}^{2}\right] +\int dA\left[\frac{1}{2}\beta_{10}{\dot{\phi}}^{2}+\frac{1}{2}\beta_{11} (\bm{v}_{\tau}^{\mathrm slip})^{2}\right].
\end{align}

Minimization of ${\cal R}$ gives the following simplified dynamic equation system: 

(i) The dynamic equations for $\bm{v}$: the incompressibility condition $\nabla \cdot \bm{v}=0$, the generalized Stokes' equation in Eq.~(\ref{eq:Appl-ActivePolar-Dyn1a}) with the stress tensors $\bm{\sigma}^{\mathrm{e}}$, 
$\bm{\sigma}^{\mathrm{v}}$, and $\bm{\sigma}^{\mathrm{a}}$, given in Eqs.~(\ref{eq:Appl-ActivePolar-Sigmae}),
(\ref{eq:Appl-ActivePolar-Sigmav}), and (\ref{eq:Appl-ActivePolar-sigmaa}), respectively; 

(ii) The dynamic equation for $\phi$: the conservation equation in Eq.~(\ref{eq:Appl-ActivePolar-phit}) with $\bm{J}=-M\nabla \mu_{\phi}$;

(iii) The dynamic equations for $\bm{p}$:
\begin{equation}\label{eq:Appl-ActivePolar-Dyn2b}  
\bm{P}=\gamma_{1}^{-1}\bm{h}-\gamma_{1}^{-1}\gamma_2\bm{p}\cdot\dot{\bm{\epsilon}}+\gamma_{1}^{-1}\bm{h}^{\mathrm{a}},
\end{equation}   
with the active molecular field $\bm{h}^{\mathrm{a}}$ given in Eq.~(\ref{eq:Appl-ActivePolar-ha2}) and the passive molecular field $h$ given in Eq.~(\ref{eq:Appl-ActivePolar-h});

(iv) The boundary conditions in Eqs.~(\ref{eq:Appl-ActivePolar-BCs1}) still apply to the present case. 

\textcolor{black}{We would like to point out that in Sec.~\ref{Sec:Appl-ActivePolar-Reactive}, 
a complete model is constructed to incorporate the chemical reaction and explicitly describe the mechanochemical coupling. 
In this description, the time-reversal symmetry (TRS) is preserved, and so is Onsager’s reciprocal relation (ORR) for mechanochemical coupling. 
In Sec.~\ref{Sec:Appl-ActivePolar-Force}, the limit of constant reaction rate is taken, and a simplified model is obtained from the complete one 
in Sec.~\ref{Sec:Appl-ActivePolar-Reactive}. In this limit, the TRS is lost, and so is ORR for mechanochemical coupling. 
However, the Parodi relation for the cross coupling in the passive polar fluid is still preserved.}

Finally, we note that the dynamic equation system for two-phase active polar flows on solid substrates is similar to that for two-phase passive polar flows on solid substrates, but is supplemented by some extra active terms (here the active stress $\bm{\sigma}^{\mathrm{a}}$ and the active molecular field $\bm{h}^{\mathrm{a}}$) that break the TRS. This type of diffuse-interface model has been solved numerically as a minimal model for cell motility~\cite{Cates2012,Cates2015}. 


%
%
%

\section{Applications 3: Dynamics of thin active  droplets on solid substrates}
\label{Sec:Appl-DropSpread}
In this section, we consider a thin droplet of active polar fluids moving on a solid substrate in a simple two dimensions ($xz$ plane), as schematically shown in Fig.~\ref{Fig:ActiveDrop}b. The detailed dynamics of such a droplet can be described by the full dynamic equation system that is derived in the previous Sec. \ref{Sec:Appl-ActivePolar}. However, here we present a thin film description of the active droplet on solid substrates where the lubrication approximation applies~\cite{Ramaswamy2008,Joanny2012,Liverpool2019,Thiele2020,Ren2010}. In this case, the shape of the droplet is described by its thickness $h\left(x, t\right)$ in the vertical $z$ direction as a function of horizontal position $x$ at time $t$, as shown in Fig.~\ref{Fig:ActiveDrop}b. The hydrodynamic velocity field and the average orientation of the active polar filaments inside the two-dimensional drop are represented by $\bm{v}(x,z,t)=\left(u, w\right)$, and the polarization vector $\bm{p}(x,z,t)$, respectively. 
\textcolor{black}{Similar problems have been studied 
by several groups~\cite{Cummings2001,Mechkov2009,Lin2013} for thin films of passive liquid crystals, 
by Kitavtsev \emph{et al.}~\cite{Kitavtsev2018} for thin films of active liquid crystals}, 
by Sankararaman \& Ramaswamy~\cite{Ramaswamy2008} for thin films of active polar fluids, 
and by Joanny \& Ramaswamy~\cite{Joanny2012} for thin drops of active polar fluids. 
Here we use OVP to provide an alternative derivation of thin film equations for active polar droplets 
and to find approximate scaling laws for the spreading or dewetting of the droplet on solid substrate.

\subsection{Thin film equations for active polar fluids}
\label{Sec:Appl-DropSpread-ThinEqn}

We firstly use OVP to derive the thin film equations for a thin active polar droplet moving on a solid substrate. To be specific and simple, we make several assumptions as follows. 


(i) The active fluid is incompressible, satisfying the  
incompressibility condition, $\nabla \cdot \bm{v}=\partial_x u+\partial_z w=0$, from which we get the local mass conservation equation for the evolution of film height as
\begin{equation}\label{eq:Appl-DropSpread-hconserv} 
\partial_{t} h+\partial_x \int_{0}^{h} u dz=0.
\end{equation}
Here and in the following, we use $\partial_{\alpha}$ to denote the partial derivatives with respect to the variable $\alpha$ such as time $t$, coordinates $x$ and $z$.  

(ii) The lubrication approximation~\cite{Xu2015,Doi2016,Haiqin2018,Craster2009} is applied to the thin-film dynamics of the active polar droplet on the solid substrate. 
In the long-wave limit, the characteristic film thickness $h_0$ is much smaller than the length scale $R_0$ 
for variations in the $x$ direction, \emph{i.e.}, $h_0/R_0\ll 1$. It follows that the film thickness varies slowly in space 
with $|\partial_x h| \ll 1$. Given $h_0/R_0\ll 1$ and $\nabla \cdot \bm{v}=0$, we obtain that the flow velocity $\bm{v}$ is approximately along the $x$ direction with $w\ll u$. 

(iii) The equilibrium contact angle of the droplet $\theta_e$ is very small such that Young's equation is approximated as
\begin{equation}\label{eq:Appl-DropSpread-YoungEqn}
\cos \theta_e = \frac{\gamma_{SG}-\gamma_{SL}}{\gamma} \approx 1-\frac{1}{2} \theta_e^2,
\end{equation}
where $\gamma$ is fluid-gas surface tension, $\gamma_{SG}$ solid-gas surface tension, and $\gamma_{SL}$ solid-fluid surface tension.

(iv) We only consider droplet dynamics with left-right symmetry and the droplet shape is mainly determined by its interfacial energy and the effects of nematic elastic energy can be neglected. This arises when the characteristic thickness of the droplet is much larger than $h_K\sim K_p/\gamma$ with $K_p$ being the elastic constant defined in Eq.~(\ref{eq:Appl-ActivePolar-Fp}). Then the total energy functional of the droplet is given by
\begin{align}\label{eq:Appl-DropSpread-Ftot}
{\cal F}[h,R]&\approx 2\gamma\int_0^R d x\left[1+\frac{1}{2}\left(\partial_{x} h\right)^{2}\right]+2(\gamma_{\mathrm{SL}}-\gamma_{\mathrm{SG}})R, \nonumber \\
&\approx \gamma\int_0^R d x\left[\theta_e^2+\left(\partial_{x} h\right)^{2}\right],
\end{align}
where $R$ is half of the contact length of the droplet with the solid substrate. The rate of change of the total energy is easily obtained as
\begin{equation}
\dot{\cal F}[\partial_t h,\dot{R}]=\int_0^R d x\left[-2\gamma \partial_{x}^{2} h\right]\partial_{t} h+\gamma\left[\theta_e^2-\left(\partial_{x} h\right)^{2}\right]_{x=R} \dot{R},   
\end{equation} 
where we have used the identity $\partial_t h+\dot{R}\partial_x h=0$ (obtained from $h(x=R(t),t)=0$) at the contact line $x=R$.

(v) The active filaments inside the droplet lack a head-tail polarity, that is, $\bm{p}$ and $-\bm{p}$ are equivalent, but they can show average nematic alignment. Furthermore, in the case of thin active droplets, the $z$ dependence of $\bm{p}=(p_x,p_{z})$ is determined by the equilibrium equations,
\begin{equation}\label{eq:Appl-DropSpread-pxpzEqn}
\partial_z^2p_x=\partial_z^2p_z=0,    
\end{equation}
which are obtained from the minimization of nematic elastic energy, ${\cal F}_p=\int dxdz\frac{1}{2}K_p\left[(\partial_z p_x)^2+(\partial_zp_z)^2\right]$, similarly as defined in Eq.~(\ref{eq:Appl-ActivePolar-Fp}) under the approximation of one elastic constant, $K_p$, and fixed magnitude of $\bm{p}$.

\textcolor{black}{(vi) We assume the planar anchoring conditions at any bounding surfaces with which the active filaments are in contact, that is, the polarization vector $\bm{p}$ is parallel to the tangent direction of all the bounding surfaces. Here we then have: $\bm{p}=\hat{\bm{x}}$ at $z=0$ and $\bm{p}=\hat{\bm{\tau}}\approx(1,\partial_x h)$ at the free surface $z=h$, in which $\hat{\bm{x}}$ is the unit vector along the x-direction and $\hat{\bm{\tau}}$ is the unit tangent vector of the free surface of the droplet. 
This anchoring boundary condition is mainly motivated by the stress-fiber structure in adherent cells~\cite{Sam2013a} and by the experimental observations~\cite{Kemkemer2000} on thin films of amoeboid cells, in which the cells lie in the plane of the glass slide on which they spread and form nematic liquid-crystal structures. The planar anchoring conditions at the free surface have been employed in many previous works~\cite{Liverpool2019,Joanny2012,Ramaswamy2008,Liverpool2020}. In contrast, in our diffuse-interface model of active polar droplets in Sec.~\ref{Sec:Appl-ActivePolar}, we have assumed planar anchoring condition at the solid surface but perpendicular anchoring condition at the free interface, which mimics the orientation of actin filaments in the lamellipodium of migrating cells. Such anchoring boundary conditions have also been used in many previous works~\cite{Cates2018,Cates2015,Cates2012,Thiele2020}.} 

\textcolor{black}{Then based on the assumptions in (v) and the planar anchoring conditions in (vi), we can solve the polarization vector $\bm{p}$ from Eq.~(\ref{eq:Appl-DropSpread-pxpzEqn}) independently of the flow velocity $v$ for a given drop profile $h(x,t)$ and obtain~\cite{Ramaswamy2008,Joanny2012}
\begin{equation}\label{eq:Appl-DropSpread-pxpz}
p_{x} \approx 1, \quad p_{z}\approx (z/h) \partial_{x} h. 
\end{equation}}

(vii) We consider only the dynamic limit of active polar fluids discussed in Sec.~\ref{Sec:Appl-ActivePolar-Force}, at which the effects of polarization and flow on the ATP hydrolysis are negligibly small, and the rate $r\approx \Lambda^{-1}\Delta \mu$ is a constant parameter. In this case, the active stress $\bm{\sigma}^{\mathrm{a}}=-\zeta\bm{p}\bm{p}$ in Eq.~(\ref{eq:Appl-ActivePolar-sigmaa}) is only a function of the local polarization vector $\bm{p}$ and it breaks the TRS, driving the system out of equilibrium locally. The work power done by $\bm{\sigma}^{\mathrm{a}}$ on the thin droplet is approximated to the leading order as
\begin{equation}\label{eq:Appl-DropSpread-Wa} 
\dot{\cal W}_a[u] =\int_0^R dx \int_0^h dz \left[2\zeta p_x p_z\partial_z u+2\zeta p_x^2\partial_x u\right].
\end{equation} 

(viii) Using the lubrication approximation, the dissipation functional is given to the leading order by~\cite{Xu2015,Doi2016,Haiqin2018,Craster2009}
\begin{equation}\label{eq:Appl-DropSpread-PhiF}
\Phi[u,\dot{R}]=\int_0^R dx \int_{0}^{h} dz\left[ \eta \left(\partial_{z} u\right)^{2}\right]+\zeta_{\mathrm{cl}}\dot{R}^2. 
\end{equation} 
Here we have neglected the dissipation from the fluid slip at the solid surface (away from the contact line).
The first term in $\Phi$ represents the usual hydrodynamic viscous dissipation under the lubrication approximation with $\eta$ being the shear viscosity of the fluid. The second term represents the extra energy dissipation associated with the dynamics near the contact line region~\cite{Doi2016,Haiqin2018}, which can be very complex for the suspension droplets of active filaments~\cite{Thiele2020}. The phenomenological parameter $\zeta_{\mathrm{cl}}$ is the frictional coefficient at the contact line, which is infinitely large for a pinned contact line, and is zero for a freely moving contact line.   

From the above discussions, we then obtain the Rayleighian as
\begin{equation}\label{eq:Appl-DropSpread-Ray}
{\cal R}[u,w,\partial_t h,\dot{R}]= \dot{\mathcal{F}}[\partial_t h,\dot{R}]+ {\Phi}[u,\dot{R}]-{\dot{\cal W}_a}[u]-\int dx dz \left[p\left(\partial_x u+\partial_z w\right)\right],
\end{equation}
where $p$ is the pressure (a Lagrange multiplier) that imposes the incompressibility constraint. 
Minimizing ${\cal R}$ with respect to the rates, $u$, $w$, $\partial_{t} h$, and $\dot{R}$, give a closed dynamic equation system for the thin droplets moving on solid substrates as follows. In the bulk, the dynamic equations are
\begin{subequations}\label{eq:Appl-DropSpread-DynBulk}
\begin{equation}\label{eq:Appl-DropSpread-DynBulk1}
\eta \partial_{z}^{2} u -\partial_x p-\frac{\zeta}{h}\partial_x h =0,
\end{equation} 
\begin{equation}\label{eq:Appl-DropSpread-DynBulk2}
\partial_{z} p=0,    
\end{equation} 
\end{subequations}
which are supplemented with the no-slip condition $u=0$ at the solid surface $z=0$, the boundary conditions at the free surface $z=h$:
\begin{subequations}\label{eq:Appl-DropSpread-DynBCzh}
\begin{equation} \label{eq:Appl-DropSpread-DynBCzh1}
p(x,h,t)=p_0-\gamma \partial_{x}^{2} h,  
\end{equation} 
\begin{equation}\label{eq:Appl-DropSpread-DynBCzh2} 
\eta \partial_{z} u|_{z=h}=0,
\end{equation} 
\end{subequations}
and the boundary conditions at the contact line $x=R$ and $z=0$:  
\begin{equation}\label{eq:Appl-DropSpread-DynMCL}
\zeta_{\mathrm{cl}} \dot{R}=\frac{\gamma}{2}\left[\left(\partial_{x} h\right)^{2}-\theta_e^2\right]. 
\end{equation} 
Here $p_0$ is the pressure in the surrounding gas; we have also used the impermeability condition $w=0$ at the solid surface $z=0$, and the kinematic boundary condition $w=\partial_{t} h+u \partial_{x} h$ at the free surface $z=h$.

As in the classical problems of thin film fluids~\cite{Oron1997,Xu2015,Ren2010}, the solution of the above closed equation system gives a parabolic profile for $u(x, z, t)$ in the form of
\begin{equation}\label{eq:Appl-DropSpread-uxzt}
u(x, z, t)=\left(-\frac{\gamma}{\eta} \partial_{x}^3 h+\frac{\zeta}{\eta h} \partial_{x} h \right)\left(\frac{z^{2}}{2}-z h\right),
\end{equation}   
from which we find the thickness-averaged velocity $\bar{u}\equiv \frac{1}{h}\int_{0}^{h} dz u$ as
\begin{equation}\label{eq:Appl-DropSpread-ubar}
\bar{u}=\frac{h^{2}}{3\eta} \left(\gamma \partial_{x}^{3} h-\zeta {h}^{-1} \partial_{x} h\right).
\end{equation} 
Substituting Eq.~(\ref{eq:Appl-DropSpread-ubar}) into the mass conservation equation (\ref{eq:Appl-DropSpread-hconserv}), one obtains the thin film equation for active droplets:
\begin{equation}\label{eq:Appl-DropSpread-ThinFilmEqn}
\partial_{t} h+\frac{1}{3 \eta} \partial_{x}\left(\gamma h^{3} \partial_{x}^3 h-\zeta h^{2} \partial_{x} h\right)=0,
\end{equation}
which is supplemented with the boundary condition (\ref{eq:Appl-DropSpread-DynMCL}) at the contact line $x=R(t)$. 
The static solutions of this thin film equation yield the steady-state shape of the active droplet and the dynamic scaling properties of the solutions lead to the spreading or dewetting laws for the active droplet~\cite{Joanny1992,DeGennes1985}. 

\textcolor{black}{We would like to comment and compare our model for the thin active droplets on solid substrates with other models in the literature~\cite{Joanny2012,Liverpool2019,Thiele2020} as follows. }

\textcolor{black}{(i) In comparison to the thin-film model by Joanny \& Ramaswamy~\cite{Joanny2012}, we have neglected the effect of nematic elastic energy in determining the droplet shape. As a result, our thin film equation is a limiting case of their model when the droplet thickness is much larger than $h_K\sim K_p/\gamma$ as discussed above near Eq.~(\ref{eq:Appl-DropSpread-Ftot}). However, if $h_K$ is not very small, our boundary condition (\ref{eq:Appl-DropSpread-DynMCL}) in the vicinity of the contact line with $h_K>h\sim 0$ may be problematic and elastic contributions have to be included.}

\textcolor{black}{(ii) In the thin-film model by Loisy \emph{et al.}~\cite{Liverpool2019}, the flows inside the active droplets are induced by the winding of the polarization field. This winding introduces a dramatic change in the orientation of the polarization $\bm{p}$ along the thickness $z$-direction: $\theta \approx \frac{z}{h}\frac{\omega \pi}{2}$, in which $\theta$ is the angle of $\bm{p}$ relative to the $x$-axis and $\omega$ is an integer winding number that counts the number of quarter turns of $\bm{p}$ across the drop height. In comparison, in the present work, we have not considered such internal polarization winding and the orientation of $\bm{p}$ varies along $z$ as $\theta \sim p_z/p_x \approx (z/h) \partial_x h$ (see Eq.(\ref{eq:Appl-DropSpread-pxpz})). Such difference in the variation of $\bm{p}$ orientation leads to the difference in the final form of thin film equation between the present work and that by Loisy \emph{et al.}~\cite{Liverpool2019}.}

\textcolor{black}{(iii) In the recent work by Trinschek \emph{et al.}~\cite{Thiele2020}, the authors have proposed a more complete model for active polar droplets, which is similar to our model presented in Sec.~\ref{Sec:Appl-ActivePolar} but have introduced one additional active contribution from the treadmilling or self-propulsion of active units in the direction of their polarization. Using their more complete free energy, we can still apply our variational approach and the lubrication approximation to study the more complicated thin film dynamics by following similar methods that we have done for thin films of binary mixtures before~\cite{Xu2015}.}

\subsection{Spreading laws for thin active fluid droplets}\label{Sec:Appl-DropSpread-SpreadLaw}

Now we consider the spreading dynamics of thin active droplets on solid substrates. We do not try to solve the thin film equation (\ref{eq:Appl-DropSpread-ThinFilmEqn}) directly, but use OVP as an approximation method to solve the scaling laws for the droplet spreading. 

We assume that the height profile of the droplet $h(x,t)$ is given by a parabolic function
\begin{subequations}\label{eq:Appl-DropSpread-SpreadLaw-profile}
\begin{equation}\label{eq:Appl-DropSpread-SpreadLaw-h}
h(x,t)= \frac{1}{2} \theta(t) R(t) \left(1-\frac{x^2}{R^2} \right),
\end{equation}
and the velocity $u(x,z,t)$ inside the droplet takes the following parabolic profile along the $z$-direction as
\begin{equation}\label{eq:Appl-DropSpread-SpreadLaw-u}
u(x,z,t)= -u_0(x,t) \left(\frac{z^2}{2h^2}-\frac{z}{h} \right).
\end{equation}
\end{subequations}
The reason of choosing these function forms is apparent  if we compare Eq.~(\ref{eq:Appl-DropSpread-SpreadLaw-u}) with the velocity profile in Eq.~(\ref{eq:Appl-DropSpread-uxzt}), and remember that the equilibrium shape of a thin droplet takes the form of Eq.~(\ref{eq:Appl-DropSpread-SpreadLaw-h}) with $\theta(t)=\theta_e$ being the equilibrium contact angle and $R(t)=R_e$ being (half of) the contact length between the droplet and the solid substrate at equilibrium. Note that $\theta(t)$ defined from the parabolic height profile in Eq.~(\ref{eq:Appl-DropSpread-SpreadLaw-h}) is the apparent contact angle interpolated away from the microscopic contact line, which is different from the  contact angle $\theta=|\partial_x h|$ defined in the previous subsection locally in the close vicinity of the contact line.

To achieve an approximate description of the droplet dynamics, the time-dependent parameters $\theta(t)$, $R(t)$, and $u_0(x,t)$ must be determined by OVP. However, note that these parameters are not independent. Firstly, from the conservation of the droplet area (or mass), $A_0$, that is, $A_0=2\int_0^{R(t)} dx h(x,t)=\mathrm{const.}$, we have
\begin{equation}\label{eq:Appl-DropSpread-SpreadLaw-AreaConserv}
\frac{2}{3} \theta(t) R(t)^2 =A_0.
\end{equation}
Secondly, from the mass conservation equation (\ref{eq:Appl-DropSpread-hconserv}) for droplet height, we obtain
\begin{equation}\label{eq:Appl-DropSpread-SpreadLaw-ubar}
\bar{u}\equiv \frac{1}{h} \int_0^h dz u(x,z,t) = -\frac{1}{h} \frac{d}{dt} \int_0^x dx' h(x',t).
\end{equation}
Substituting the approximate profile of $h(x,t)$ and $u(x,z,t)$ in Eqs.~(\ref{eq:Appl-DropSpread-SpreadLaw-profile}) into Eq.~(\ref{eq:Appl-DropSpread-SpreadLaw-ubar}) we obtain
\begin{equation}
u_0(x,t)=3x\frac{\dot{R}(t)}{R(t)}. 
\end{equation}
Therefore, the dynamics of the droplet can be described by one time-dependent parameter and here we take $R(t)$.

Substituting the droplet profile $h(x,t)$ in Eq.~(\ref{eq:Appl-DropSpread-SpreadLaw-h}) into Eq.~(\ref{eq:Appl-DropSpread-Ftot}), we obtain the total free energy
\begin{equation}\label{eq:Appl-DropSpread-SpreadLaw-Ftot}
{\cal F}[h,R] = \gamma\left(\theta_e^2R+\frac{3A_0^2}{4R^3}\right),
\end{equation}
from which we find the rate of change of the total energy as
\begin{equation}\label{eq:Appl-DropSpread-SpreadLaw-Fdot}
\dot{\cal F}[\dot{R}] = \gamma\left(\theta_e^2-\frac{9A_0^2}{4R^4}\right)\dot{R}.
\end{equation}
Similarly, we find the energy dissipation function $\Phi$ as 
\begin{equation}\label{eq:Appl-DropSpread-SpreadLaw-PhiF}
\Phi[\dot{R}] = \frac{C\eta R^{2}}{A_0} \dot{R}^{2}+\zeta_{\mathrm{cl}}\dot{R}^2,
\end{equation}
and the work power done by active stress, $\bm{\sigma}^{\mathrm{a}}$, as
\begin{equation}\label{eq:Appl-DropSpread-SpreadLaw-Wa} 
\dot{\cal W}_a[\dot{R}] = \frac{\zeta A_0}{R}\dot{R},
\end{equation}
which are obtained by substituting the approximate profile of $h(x,t)$ and $u(x,z,t)$ in Eqs.~(\ref{eq:Appl-DropSpread-SpreadLaw-profile}) into the dissipation function in Eq.~(\ref{eq:Appl-DropSpread-PhiF}) and the active work power function in Eq.~(\ref{eq:Appl-DropSpread-Wa}), respectively. Here $C\equiv 2\left[\ln \left({2R}/{\epsilon}\right)-2\right]$ and $\epsilon$ is the molecular cutoff length that is introduced to remove the divergence in the energy dissipation near the contact line.

Then from Eqs.~(\ref{eq:Appl-DropSpread-SpreadLaw-Fdot})--(\ref{eq:Appl-DropSpread-SpreadLaw-Wa}), we obtain the Rayleighian ${\cal R}[\dot{R}]=\dot{\mathcal{F}}+ {\Phi}-{\dot{\cal W}_a}$. 
Minimizing ${\cal R}$ with respect to $\dot{R}$ gives the following evolution equation 
\begin{equation}\label{eq:Appl-DropSpread-SpreadLaw-REqn}
\left(1+k_{\mathrm{cl}}\right) \tau_{\mathrm{act}} \dot{R}=\mathrm{sign}(\zeta)\frac{A_{0}^{2}}{2CR^3}+\frac{A_{0}^{\frac{1}{2}} \theta\left(\theta^{2}-\theta_{e}^{2}\right)}{3 C k_{\mathrm{act}} \theta_{e}^{3}}.
\end{equation}
Here $\mathrm{sign}(\zeta)$ is the sign function of $\zeta$ with $\mathrm{sign}(\zeta)=+1$ for extensile active units and $\mathrm{sign}(\zeta)=-1$ for contractile active units as shown in Fig.~\ref{Fig:forcedipoles}. The contact angle $\theta$ is a function of $R$ as given by Eq.~(\ref{eq:Appl-DropSpread-SpreadLaw-AreaConserv}). The dimensionless parameter $k_{\mathrm{cl}} \equiv \zeta_{\mathrm{cl}} / \zeta_{\mathrm{hydro}}$ is a material parameter determined by the droplet and the substrate, which can be treated as a constant and characterizes the importance of the additional friction $\zeta_{\mathrm{cl}}$ near the contact line relative to the normal hydrodynamic friction $\zeta_{\mathrm{hydro}}=3C\eta/2\theta$ in the bulk fluids of the droplet. Two time scales are introduced and defined as
\begin{equation}\label{eq:Appl-DropSpread-SpreadLaw-TwoTimes}
\tau_{\mathrm{act}}\equiv \frac{\eta}{|\zeta|}, \quad \tau_{\mathrm{rel}} \equiv \frac{\eta A_0^{1/2}}{\gamma \theta_e^3}.
\end{equation} 
The time $\tau_{\mathrm{act}}$ represents the characteristic time for the droplet to reach steady-state motion driven by active stresses, and $\tau_{\mathrm{rel}}$ represents the relaxation time needed for the droplet to reach the equilibrium contact angle $\theta_{e}$.   
The dimensionless parameter $k_{\mathrm{act}}$ defined by
\begin{equation}\label{eq:Appl-DropSpread-SpreadLaw-kact}
k_{\mathrm{act}}\equiv \frac{\tau_{\mathrm{rel}}}{\tau_{\mathrm{act}}}    
\end{equation}
characterizes the strength of activity in the active fluids. If  $k_{\mathrm{act}}$ is small, the equilibration of the droplet shape and contact angle is very fast and the droplet spreading is mainly driven by surface energy. On the other hand, if $k_{\mathrm{act}}$ is large, the activity is strong and the activity-driven droplet motion is much faster than the energy-driven equilibration of the droplet, that is, the droplet spreading is mainly driven by active stress and a large Capillary-number flow will be induced. 

Particularly, for $k_{\mathrm{act}} \ll 1$, the first term on the right-hand side of Eq.~(\ref{eq:Appl-DropSpread-SpreadLaw-REqn}) can be ignored, and the evolution equation for $R$ becomes
\begin{equation}\label{eq:Appl-DropSpread-SpreadLaw-REqn1}
\left(1+k_{\mathrm{cl}}\right) \tau_{\mathrm{rel}} \dot{R}=\frac{A_{0}^{\frac{1}{2}} \theta\left(\theta^{2}-\theta_{e}^{2}\right)}{3 C \theta_{e}^{3}},
\end{equation}
which gives (for $\theta_e \to 0$) the classical Tanner's spreading law for two-dimensional droplets~\cite{Joanny1992,DeGennes1985,Bonn2009}:
\begin{equation}\label{eq:Appl-DropSpread-SpreadLaw-Law1}
R(t)\sim A_0^{1/2}\left(\frac{t}{\tau_{\mathrm{rel}}}\right)^{1/7}=\left(\frac{\gamma A_0^3 t}{\eta}\right)^{1/7}. 
\end{equation}
On the other hand, if $k_{\mathrm{act}} \gg 1$, the second term on the right-hand side of Eq.~(\ref{eq:Appl-DropSpread-SpreadLaw-REqn}) can be ignored, and the equation becomes
\begin{equation}\label{eq:Appl-DropSpread-SpreadLaw-REqn2}
\left(1+k_{\mathrm{cl}}\right) \tau_{\mathrm{act}} \dot{R}=\mathrm{sign}(\zeta)\frac{A_{0}^{2}}{2CR^3},
\end{equation}
from which we obtain the spreading law (or dewetting law from very small initial contact angle) predicted by Joanny \& Ramaswamy~\cite{Joanny2012} for two-dimensional droplets of extensile active units with $\mathrm{sign}(\zeta)=+1$ (or of contractile active units with $\mathrm{sign}(\zeta)=-1$):
\begin{equation}\label{eq:Appl-DropSpread-SpreadLaw-Law2}
R(t)\sim A_0^{1/2}\left(\frac{t}{\tau_{\mathrm{act}}}\right)^{1/4}=\left(\frac{|\zeta|A_0^2 t}{\eta}\right)^{1/4}. 
\end{equation}

Note that as mentioned by Joanny \& Ramaswamy~\cite{Joanny2012}, the effects of activity on the droplet spreading enter at the same order in gradients as those of gravity, but with a different dependence on the film height. Furthermore, similar dynamic equation as Eq.~(\ref{eq:Appl-DropSpread-SpreadLaw-REqn}) for thin droplets on solid substrates has been obtained in a very different scenario where evaporation occurs at the free surface of the droplet~\cite{Doi2016}. The effects of activity on the droplet spreading enter at the same order as those of evaporation, but with a different dependence on the droplet radius or contact length. 

In addition, the formulation and calculations presented here can be readily extended to the thin-film dynamics of three-dimensional droplets on solid substrates, particularly for the spreading dynamics of a droplet with cylindrical symmetry. Furthermore, the effects of nematic energy on the spreading dynamics can also be considered by including nematic elastic energy~\cite{Joanny2012}, which takes the simple form of ${\cal F}_p=\int dx \frac{1}{2}K_p\left[(\partial_xh)^2/h\right]$. 






\section{Conclusions}
\label{Sec:Conc}
Onsager’s variational principle (OVP) has recently become an indispensable and powerful tool in the study of the nonlinear and nonequilibrium phenomena of many inert soft matter systems, such as liquid droplets, colloid suspensions, nematic liquid crystals, polymer gels, and surfactants, \emph{etc}.  
In this work, we present a simple extension of OVP for the dynamic modeling of active soft matter such as suspensions of bacteria and aggregates of animal cells. In this extended OVP, the active forces generated locally by individual active units are treated as non-conservative forces that cannot be derived from any free energy and dissipation functions. We then apply this extended form of OVP to three representative active matter problems, which are motivated by the biology of bacteria and animal cells.
We show that OVP can not only help to formulate thermodynamically-consistent models, but can also be used to find approximate solutions for the emergent structures and complex dynamics of active soft matter. 

The first application of OVP presented in Sec.~\ref{Sec:Appl-ActiveUnit} is about the directional motion of individual active units: a molecular motor walking on a stiff biofilament and a toy two-sphere microswimmer moving in a viscous fluid. In the motor/filament system, we consider the  mechanochemical cross-coupling which  indicates that the system is a reversible machine: it can not only convert chemical energy into mechanical work, but can also convert mechanical work into chemical energy. In the toy microswimmer, we show how directional self-propulsion can be generated by cyclic body-shape oscillations together with front-back asymmetry in hydrodynamic friction. It is shown that mechanochemical cross-coupling in biological systems can be considered in Onsager's framework of non-equilibrium thermodynamics. Activity and the broken time reversal symmetry in active matter are basically resulted from the persistent consumption and conversion of chemical energy, released during spontaneous ATP hydrolysis, into motion or mechanical work.  

The second application presented in Sec.~\ref{Sec:Appl-ActivePolar} is about the two-phase hydrodynamics for a droplet of active polar fluids, which is composed of suspending contractile or extensile active units such as bacteria, actomyosin units, and animal cells. We use OVP to formulate a diffuse-interface model for an active polar droplet moving on a solid substrate. This hydrodynamic model is thermodynamically consistent in both hydrodynamic equations in the bulk fluid and matching boundary conditions at the solid surface.  

The third application presented in Sec.~\ref{Sec:Appl-DropSpread} is about the motion of a thin active polar droplet on a solid substrate in two dimensions. Using the lubrication approximation, we firstly apply OVP to derive the classical thin film equation that has been obtained previously. 
We then use OVP as an approximation tool to find two scaling laws for the spreading (or dewetting) of the thin active droplet in the respective limits of negligible activity and strong activity. It is interesting to note that the reduced equation obtained for the spreading (or dewetting) dynamics of thin active droplets takes a similar form as that has been obtained previously for the dewetting dynamics of an evaporating droplet on solid substrates. 


Below we make a few general remarks and outlook.

(i) \emph{Near-equilibrium assumption of OVP.} OVP is proposed in Onsager's linear-response framework of non-equilibrium thermodynamics, which is based on the near-equilibrium assumption. However, biological systems are usually far away from equilibrium. Therefore, the validity and the range of the OVP applications should be and can only be justified by solving real biological problems and comparing with quantitative experiments~\cite{Marchetti2013,Prost2015}. 

\textcolor{black}{(ii) \emph{Relationships between OVP/OMVP and other approaches in nonequilibrium thermodynamics.} 
Following the pioneering works of Onsager, there have been various approaches 
developed for nonequilibrium thermodynamics. In particular, there have been various variational principles 
formulated for the study of irreversible processes~\cite{Ichiyanagi1994}. 
More recently, the general equation for non-equilibrium reversible-irreversible coupling (GENERIC) formalism 
has been proposed as an extension of Hamiltonian's formalism of classical mechanics to nonequilibrium thermodynamic systems 
with both reversible and irreversible dynamics~\cite{Ottinger2005}. 
However, a general discussion on the relationships among the various approaches is beyond the scope of this work.}

(iii) \emph{Applications of Onsager-Machlup variational principle (OMVP).} 
\textcolor{black}{In the end of Sec.~\ref{Sec:OVP-OVPOMVP}, we have mentioned that Onsager and Machlup~\cite{Onsager1953} introduced OMVP 
in their study of the statistical fluctuations of kinetic paths in the framework of Langevin equation. 
They have shown that the most probable kinetic path over a certain long-time period is determined by the minimization of a time integral, \emph{i.e.}, 
the Onsager-Machlup integral in Eq.~(\ref{eq:OVP-Olong-LeastSquare}). 
Recently, Doi \emph{et al.}~\cite{Doi2019} proposed that OMVP can be used to approximate the long-time dynamics of nonequilibrium systems. 
However, in this work, we have not given applications of using OMVP to find approximate solutions for long-time behaviors such as steady states. 
We mention two potential applications of OMVP as follows: the steady-state for the Wave propagation and sustained oscillations observed in migrating cells~\cite{Gerisch2004,Weiner2007,Inagaki2017}; the steady-state (spontaneous) retraction dynamics of an injured axon~\cite{Lin2019} or a laser-cutting stress-fiber bundle in adherent cells~\cite{Schwarz2011}.}

(iv) \emph{Applications of OVP and OMVP to more specific biological problems.} The applications considered in this work are mostly toy models or simplified models of mostly theoretical interests. We are now trying to apply the extended form of OVP to more specific biological problems such as cell spreading, cell curvotaxis, wound closure, tissue folding, and so on. However, in these real systems, we usually need to involve many more complex active processes~\cite{Julicher2017} in addition to active forces and cyclic body-shape oscillations, such as tensional homeostasis~\cite{Sam2013a}, cell division and apoptosis~\cite{Joanny2010}, topological cell rearrangements~\cite{Julicher2017}, memory effects~\cite{Marchetti2013,Marchetti2019}.


In summary, the variational method proposed in this work about incorporating biochemical activity into OVP will help to construct thermodynamically-consistent models and to find approximate dynamic solutions in active soft matter. Particularly, this will help to deepen our understanding of the emergent structure and dynamic behaviors of real \emph{in vivo} biological systems such as bacteria suspensions, individual animal cells and cell aggregates (or tissues)~\cite{Joanny2010,Sam2013a,Marchetti2013}.   


\appendix

\section*{Appendices} 

\addcontentsline{toc}{section}{Appendices}
\renewcommand{\thesubsection}{\Alph{subsection}} 
\renewcommand{\theequation}{\Alph{subsection}\arabic{equation}}
\setcounter{equation}{0}  
\subsection{Some additional notes on the choice of thermodynamic fluxes and forces}\label{Sec:AppA}

In Onsager's theory of non-equilibrium thermodynamics, the choice of thermodynamic fluxes and forces is not unique~\cite{deGroot1984,Gyarmati1970}. For example, for active polar fluids, two different sets of fluxes and forces can be chosen as follows:

(i) Fluxes are chosen to be $r$, $\bm{P}$, $\dot{\bm{\epsilon}}$ (or the flow velocity $\bm{v}$) and the corresponding forces are $\Delta \mu$, $\bm{h}$, $\bm{\sigma}^v$, respectively, as taken in Sec.~\ref{Sec:Appl-ActivePolar} of this work. Here all the three fluxes have the same time parity.

(ii) Fluxes are chosen to be $r$, $\bm{P}$, $\bm{\sigma}^v$ (the momentum flux) and the corresponding forces are $\Delta \mu$, $\bm{h}$, $\dot{\bm{\epsilon}}$, respectively, as taken by Marchetti \emph{et al.}~\cite{Marchetti2013}. Here the time parity of the flux $\bm{\sigma}^v$ is different from the other two fluxes $r$ and $\bm{P}$. Furthermore, note that in this case, the new pair of flux $\bm{\sigma}^v$ and force $\dot{\bm{\epsilon}}$ is a swap of the pair of flux $\dot{\bm{\epsilon}}$ and force $\bm{\sigma}^v$ in the first choice (i). 

The symmetry of Onsager matrix coupling fluxes and forces depends on the time parity (\emph{i.e.}, the time-reversal signature) of the fluxes~\cite{deGroot1984,Marchetti2013}: the Onsager coupling matrix is symmetric for fluxes of the same time parity and is antisymmetric for fluxes of opposite time parity. Therefore, for the first choice (i) with fluxes of the same time parity, the Onsager coupling matrix is symmetric. In comparison, for the second choice (ii), the cross-coupling coefficients between the flux $\bm{\sigma}^v$ and the flux $r$ or between the flux $\bm{\sigma}^v$ and the flux $\bm{P}$ are both antisymmetric as shown in Marchetti \emph{et al.}~\cite{Marchetti2013}. 


Given the non-unique choice of flux-force pairs, we will make some general discussions in this appendix about the consequences of different choices of thermodynamic fluxes and forces. 

\subsubsection{Changes in the symmetry of Onsager coupling matrix by swapping some of fluxes and forces}\label{Sec:AppA-swap}

We discuss the changes in the symmetry of Onsager coupling matrix when we swap some of fluxes with forces, as in the above two choices of force-flux pairs in active polar fluids. To this end, let's consider two pairs of fluxes and forces: $\dot{\alpha}_1$ and $X_1$, $\dot{\alpha}_2$ and $X_2$. Suppose that the two fluxes $\dot{\alpha}_1$ and $\dot{\alpha}_2$ have the same time parity, and hence the two forces $X_1$ and $X_2$ also have the same time parity but opposite to that of the corresponding fluxes. In this case, the linear force-flux relations are then given by:
\begin{equation}\label{eq:AppA-X1X2} 
\begin{array}{l}
X_1=L_{11} \dot{\alpha}_{1}+L_{12} \dot{\alpha}_{2}, \\
X_2=L_{21}  \dot{\alpha}_{1}+L_{22} \dot{\alpha}_{2},
\end{array}
\end{equation}
in which the Onsager coupling matrix $L_{ij}$ is symmetric and positive definiteness, that is, $L_{12}=L_{21}$, $L_{11},\,L_{22}>0$, and $L_{11}L_{22}-L_{12}^2>0$. 

If alternatively, we swap $X_2$ with $\dot{\alpha}_2$, that is, we choose the fluxes to be $\dot{\alpha}_1$ and $X_2$, and the corresponding forces are $X_1$ and $\dot{\alpha}_2$. Then the time parities of the two fluxes $\dot{\alpha}_1$ and $X_2$ are now different. In this case, the linear force-flux relations become
\begin{equation}\label{eq:AppA-X1X22}
\begin{array}{l}
X_1=\tilde{L}_{11}\dot{\alpha}_{1}+\tilde{L}_{12} X_2, \\
\dot{\alpha}_{2}=\tilde{L}_{21} \dot{\alpha}_{1}+\tilde{L}_{22} X_2
\end{array}
\end{equation}
in which $\tilde{L}_{11}={(L_{11}L_{22}-L_{12}^2)}/{L_{22}}$, $\tilde{L}_{12}=L_{12}/{L_{22}}$, $\tilde{L}_{21}=-L_{21}/{L_{22}}$, $\tilde{L}_{22}=1/{L_{22}}$. Note that now $\tilde{L}_{21}=-\tilde{L}_{12}$, that is, the Onsager coupling matrix $\tilde{L}_{ij}$ is antisymmetric. From this simple example, we show that although there exists some flexibility in choosing fluxes and forces, we can safely use the time parity of the chosen fluxes to determine the symmetry of the Onsager coupling matrix in the linear force-flux relations. 

\subsubsection{Onsager's variational principle for thermodynamic fluxes of different time parities}\label{Sec:AppA-TimeParity}

Suppose that there are two sets of slow state variables, $\bm{\alpha}$ and $\bm{\beta}$ and they have opposite time parity. 
Then the free energy is $\mathcal{F}(\bm{\alpha},\bm{\beta})$ and the rate of change of the free energy is given by
\begin{equation}\label{eq:AppA-Fdot}
\dot{\mathcal{F}}=-X_{i} \dot{\alpha}_{i}-Y_{i} \dot{\beta}_{i},
\end{equation}
in which $\dot{\bm{\alpha}}$ and $\dot{\bm{\beta}}$ are two fluxes, and
\begin{equation}\label{eq:AppA-fg}
X_{i} \equiv-\partial \mathcal{F} / \partial \bm{\alpha}_{i}, \quad 
Y_{i} \equiv-\partial \mathcal{F} / \partial \beta_{i}
\end{equation}
are the corresponding conjugate forces. Since $\bm{\alpha}$ and $\bm{\beta}$ have opposite time parities, then the time parities of $\dot{\bm{\alpha}}$ and $\dot{\bm{\beta}}$, and that of $\bm{X}$ and ${\bm{Y}}$ are both opposite, respectively. In this case, we can decompose the two forces into reactive and dissipative parts as
\begin{equation}\label{eq:AppA-fgdecomp}
\bm{X}=\bm{X}^{\mathrm{d}}+\bm{X}^{\mathrm{r}}, \quad \bm{Y}=\bm{Y}^{\mathrm{d}}+\bm{Y}^{\mathrm{r}}.
\end{equation}
Here $\bm{X}^{\mathrm{d}}$ and $\bm{Y}^{\mathrm{d}}$ arise from dissipative couplings that can be derived from dissipation function, while $\bm{X}^{\mathrm{r}}$ and $\bm{Y}^{\mathrm{r}}$ derive from reactive couplings:
\begin{equation}\label{eq:AppA-frgr}
X_{i}^{r}=L_{i j}^{\alpha \beta} \dot{\beta}_{j}, \quad
Y_{i}^{r}=L_{i j}^{\beta \alpha} \dot{\alpha}_{j}
\end{equation}
with the friction matrix $L_{i j}^{\alpha \beta}=-L_{j i}^{\beta \alpha}$ being antisymmetric. The reactive couplings do not contribute to dissipation and therefore can not be derived from dissipation functions. Substituting Eq.~(\ref{eq:AppA-fgdecomp}) into the rate of change of the free energy in Eq.~(\ref{eq:AppA-Fdot}), using Eq.~(\ref{eq:AppA-frgr}) and $L_{i j}^{\alpha \beta}=-L_{j i}^{\beta \alpha}$, we obtain
\begin{equation}\label{eq:AppA-Fdot2}
\dot{\mathcal{F}}=-X_{i}^{d} \dot{\alpha}_{i}-Y_{i}^{d} \dot{\beta}_{i}.
\end{equation}

The dissipation function is given by
\begin{equation}\label{eq:AppA-Phi}
\Phi=\frac{1}{2} L_{i j}^{\alpha \alpha} \dot{\alpha}_{i} \dot{\alpha}_{j}+\frac{1}{2} L_{i j}^{\beta \beta} \dot{\beta}_{i} \dot{\beta}_{j}
\end{equation}
with the friction matrix $L_{i j}^{\alpha \alpha}=L_{j i}^{\alpha \alpha}$ and $L_{i j}^{\beta \beta}=L_{j i}^{\beta \beta}$ being symmetric. Using Eqs.~(\ref{eq:AppA-fgdecomp})--(\ref{eq:AppA-Phi}) and minimizing the Rayleighian $\mathcal{R}[\dot{\bm{\alpha}}, \dot{\bm{\beta}}]=\dot{\mathcal{F}}+\Phi$, we obtain
\begin{equation}\label{eq:AppA-XY}
\begin{array}{l}
X_{i}=L_{i j}^{\alpha \alpha} \dot{\alpha}_{j}+L_{i j}^{\alpha \beta} \dot{\beta}_{j} \\
Y_{i}=L_{i j}^{\beta \alpha} \dot{\alpha}_{j}+L_{i j}^{\beta \beta} \dot{\beta}_{j} 
\end{array}
\end{equation}
or equivalently by taking the inverse:
\begin{equation}\label{eq:AppA-YX}
\begin{array}{l} 
\dot{\alpha}_{i}=R_{i j}^{\alpha \alpha} X_{j}+R_{i j}^{\alpha \beta} Y_{j} \\
\dot{\beta}_{i}=R_{i j}^{\beta \alpha} X_{j}+R_{i j}^{\beta \beta} Y_{j},
\end{array}
\end{equation}
where the mobility matrix $\bm{R}$ is the inverse of the friction matrix $\bm{L}$. Note that only the dissipative part of the thermodynamic forces can be considered and derived from the minimization of the Rayleighian, which is simply because the reactive forces don't contribute to the dissipation function and the Rayleighian. 






\section*{Conflicts of interest}
There are no conflicts to declare. 

\section*{Acknowledgements}
We thank Masao Doi at Beihang University for many fruitful discussions, particularly during his visit in Xu's group in 2019. Great thanks should also be given to Len Pismen and Yariv Kafri at Technion for their valuable comments and suggestions. T.Q. acknowledges support of the Hong Kong RGC CRF (No.~C1018-17G).
X.X. is supported in part by National Natural Science Foundation of China (NSFC, No.~12004082), by Provincial Science Foundation of Guangdong (2019A1515110809), by Guangdong Province Universities and Colleges Pearl River Scholar Funded Scheme (2019), by 2020 Li Ka Shing Foundation Cross-Disciplinary Research Grant (No.~2020LKSFG08A), by Guangdong Basic and Applied Basic Research Foundation (2020B1515310005), and by Featured Innovative Projects (No.~2018KTSCX282) and Youth Talent Innovative Platforms (No.~2018KQNCX318) in Universities in Guangdong Province.   


\balance


\bibliography{OnsagerActive} 
\bibliographystyle{rsc} 

\end{document}